\documentclass[11pt,3p,review,authoryear]{elsarticle}

\biboptions{longnamesfirst}
\usepackage[dvipsnames]{xcolor}
\usepackage{soul,framed} %,caption
\usepackage{mathtools}
\usepackage{lscape}
\colorlet{shadecolor}{yellow}
\usepackage[ruled,vlined]{algorithm2e}
\usepackage{amsmath}
\usepackage{amsfonts}
\usepackage{amssymb}
\usepackage{subcaption}
\usepackage{hyperref}
\usepackage{array}

\usepackage[utf8]{inputenc}
\usepackage{placeins}

\usepackage{blindtext}
\usepackage{dirtytalk}
\usepackage{adjustbox}
\usepackage{pdfpages} % include pdfs
\usepackage{pgffor} % for loops
\usepackage{xr}
\makeatletter
\AtBeginDocument{\let\LS@rot\@undefined}
\cleardoublepage\makeatother
\usepackage[shortlabels]{enumitem}
\usepackage{mdwtab}
\usepackage{eqparbox}
\usepackage{url}

\newtheorem{prop}{Proposition}
\newtheorem{remark}{Remark}
\usepackage{mathtools}
\usepackage{dsfont}
\usepackage{bbold}
\usepackage[english]{babel} % To obtain English text with the blindtext package
\usepackage{blindtext}
\usepackage{epstopdf}
\DeclareGraphicsExtensions{.png,.pdf,.jpg}
\usepackage{float}
\usepackage{hyperref}
\hypersetup{
    colorlinks=true,
    linkcolor=blue,
    filecolor=magenta,      
    urlcolor=cyan,
    pdftitle={Overleaf Example},
    pdfpagemode=FullScreen,
    }
\usepackage[utf8]{inputenc}
\usepackage{blindtext}
\begin{document}

\begin{frontmatter}

\title{Forecasting CPI inflation under economic policy and geopolitical uncertainties}

%% AUTHORS %%%%%%%%%%%%%%%%%%%%%%%%%%%%%%%%%%%%%%%%%%%%%%%%%%%%%%%%%%%%%%%%%%%%
%% Leave this section commented out so that the paper is blinded for review.
\author{Shovon Sengupta\textsuperscript{1,}
\footnote[0]{\textit{Equal Contributions}}, 
Tanujit Chakraborty\textsuperscript{2, 0, 3}
\footnote[5]{\textit{Corresponding author}: \textit{Mail}: tanujit.chakraborty@sorbonne.ae},
Sunny Kumar Singh\textsuperscript{1}\\
{\scriptsize \textsuperscript{1} Birla Institute of Technology And Science, Hyderabad Campus, India.}\\
{\scriptsize \textsuperscript{2} Department of Science and Engineering, Sorbonne University Abu Dhabi, UAE.} \\
{\scriptsize \textsuperscript{3} Sorbonne Center for Artificial Intelligence, Paris, France.}}

% \author[ss]{Shovon Sengupta}
% \address[ss]{}
% %\\p20200503@hyderabad.bits-pilani.ac.in
% %\\shovon.sengupta@fmr.com
% \author[tc]{Tanujit Chakraborty\corref{cor}}
% \address[tc]{Sorbonne Center for AI and Sorbonne University}
% \author[ss]{Sunny Kumar Singh}

% \cortext[cor]{Corresponding author}
% \ead{tanujit.chakraborty@sorbonne.ae}
%%%%%%%%%%%%%%%%%%%%%%%%%%%%%%%%%%%%%%%%%%%%%%%%%%%%%%%%%%%%%%%%%%%%%%%%%%%%%%%%

\begin{abstract}
Forecasting consumer price index (CPI) inflation is of paramount importance for both academics and policymakers at the central banks. This study introduces a filtered ensemble wavelet neural network (FEWNet) to forecast CPI inflation, which is tested on BRIC countries. FEWNet breaks down inflation data into high and low-frequency components using wavelets and utilizes them along with other economic factors (economic policy uncertainty and geopolitical risk) to produce forecasts. All the wavelet-transformed series and filtered exogenous variables are fed into downstream autoregressive neural networks to make the final ensemble forecast. Theoretically, we show that FEWNet reduces the empirical risk compared to fully connected autoregressive neural networks. FEWNet is more accurate than other forecasting methods and can also estimate the uncertainty in its predictions due to its capacity to effectively capture non-linearities and long-range dependencies in the data through its adaptable architecture. This makes FEWNet a valuable tool for central banks to manage inflation.

% We also demonstrate that the rolling-window real-time forecasts obtained from the proposed algorithm are significantly more accurate than benchmark forecasting methods. Additionally, we use conformal prediction intervals to quantify the uncertainty associated with the forecasts generated by the proposed approach. The excellent performance of FEWNet can be attributed to its capacity to effectively capture non-linearities and long-range dependencies in the data through its adaptable architecture.
\end{abstract}

\begin{keyword}
Inflation Forecasting \sep Wavelets \sep Neural Networks \sep Empirical Risk Minimization \sep Conformal Prediction Intervals
% Inflation forecasting \sep Macroeconomic Forecasting \sep Neural Networks \sep Conformal Prediction Intervals
% Suggested keywords are listed at https://ijf.forecasters.org/keywords/
\end{keyword}
\end{frontmatter}

\section{Introduction}
The monetary policy framework at central banks relies heavily on forecasts of various macroeconomic variables. Forecasts have also grown in importance as a means of communication for central banks during the past 20 years, and they have the potential to shape public and market expectations. Numerous studies have examined central bank forecasting from various angles due to the paramount significance of these predictions \cite{stock2007has, pratap2019rbi, faust2013forecasting, smalter2017macroeconomic, nakamura2005inflation}. Forecasting, particularly in the context of different macroeconomic policy variables, is one of the major focus areas in this domain. Inflation is one such policy variable considered instrumental in designing an economy's overall monetary policy. Thus, a deeper understanding of the factors that help forecast inflation accurately is fundamental to policymakers at the central bank \cite{stock2007has}. Over the past few decades, emerging market economies such as Brazil, Russia, India, and China (BRIC) have undergone significant changes in their macroeconomic conditions, including shifts in monetary policy frameworks such as inflation targeting, increased globalization, and the impact of events like the financial crisis and the recent pandemic. These shifts make predicting key macroeconomic variables like consumer price index (CPI) inflation quite challenging \cite{pratap2019rbi}. Given these dynamic changes in the macroeconomic environment, factors like economic policy uncertainty and geopolitical risk might offer critical insights to forecast inflation accurately.

Due to the significance of inflation forecasting exercise for monetary policy, central banks often depend on distinct models, namely structural and non-structural models. The structural models, which are different versions of Phillips curve models, are derived from economic theory. On the other hand, the non-structural models, such as univariate and multivariate time-series models, aim to utilize the unique characteristics of the data without explicitly depending on any economic theory \cite{faust2013forecasting}. Non-structural models often prioritize forecast accuracy above causal inference and are particularly valuable for short-term out-of-sample forecasting. Linear time series models such as autoregressive integrated moving average (ARIMA) and vector autoregressive (VAR) are limited in their ability to account for business cycles, periods of extreme volatility caused by macroeconomic uncertainties, and structural changes. As a result, non-linear, hybrid, and machine learning models have become increasingly popular in addressing these shortcomings. Furthermore, the existing linear time series forecasting methods bring several undesirable properties, ranging from high sensitivity to model specification to high-frequency data requirements \cite{smalter2017macroeconomic}. This is especially important for forecasting inflation or similar macroeconomic variables that are only available at the monthly, quarterly, and annual frequency. Machine learning (ML) models and combined approaches provide an opportunity to improve the forecast performance of the inflation data in emerging economies. 

Recently, an increasing number of studies have been conducted on applying ML models for inflation forecasting, both for advanced and emerging market economies. A study conducted in the United States utilized a basic feed-forward neural network to forecast quarterly CPI inflation. The study concluded that this approach was successful in predicting CPI inflation \cite{nakamura2005inflation}. Neural networks proved more effective than traditional models in predicting the monthly inflation rate for the OECD countries \cite{choudhary2012neural}. Their research revealed that neural network models outperformed simple AR models in 45\% of the nations, on average, whereas simple AR models performed better in 23\% of the countries. Recurrent neural networks, in their various forms, have been useful for predicting inflation in developed countries \cite{chen2001semiparametric, mcadam2005forecasting, almosova2023nonlinear, barkan2023forecasting}. In addition, a recent study compared various machine learning models, including lasso regression, random forests, and deep neural networks, to analyze inflation in the United States. \cite{medeiros2021forecasting} also considered several external factors, such as cash and credit availability, online prices, housing prices, consumer data, exchange rates, and interest rates \cite{medeiros2021forecasting}. Their research has shown that machine learning models with many covariates consistently outperformed conventional techniques. This superiority can be attributed to non-linear relationships between historical macroeconomic variables and inflation. This has dispelled doubts about using machine learning and contemporary deep learning architectures in analyzing big data in economics \cite{mullainathan2017machine}. For emerging market economies, the study of inflation forecasting using advanced ML models is in a nascent stage. Applications of neural networks are evident in developing countries like India, Brazil \cite{araujo2023machine}, Russia \cite{pavlov2020forecasting}, Turkey, South Africa \cite{botha2023big}, and China \cite{ozgur2021inflation}. However, none of the above studies, to the best of our knowledge, have considered exogenous factors like macroeconomic uncertainties for inflation forecasting, especially for emerging markets.

Following events such as the global financial crisis, it has been observed that many countries worldwide have encountered heightened uncertainty. This has had an impact on the decision-making process of individuals and consequently on the overall macroeconomy \cite{bloom2009impact, bloom2014fluctuations, baker2016measuring}. Although previous studies have primarily examined the connection between uncertainty and the overall economy, the evaluation of the changing relationship between measures of uncertainty and inflation is still in its early stages \cite{jones2013time, leduc2016uncertainty}. Gaining insight into the intricacies of inflation dynamics and uncertainty metrics is critical for policymakers. Multiple recent studies indicate that both economic policy uncertainty and geopolitical risk significantly influence the movement of inflation within a given sample \cite{colombo2013economic, jones2013time, balcilar2017long, benchimol2020forecast, caldara2022measuring, adeosun2023uncertainty,anderl2023asymmetries}. Consequently, inflation forecasting models that do not incorporate these uncertainties are deemed to be incorrectly specified. The study conducted by \cite{balcilar2017long} examines the effectiveness of different uncertainty measures in predicting inflation. In particular, they use US inflation data and the economic policy uncertainty index (EPU) to compare how well the vector autoregressive fractionally integrated moving average (VARFIMA) model performs at making predictions compared to other models. The study conducted by \cite{demirer2022policy} examined the predictive impact of EPU on stock market volatility in the United States and the United Kingdom. The findings suggest that EPU has a significant positive effect on volatility. However, this positive relationship weakens when the signal quality is low. \cite{liu2015economic} asserted that EPU has a crucial impact on enhancing the predictive accuracy of models forecasting stock market volatility. The study conducted by \cite{jones2016uncertain} investigated the impact of policy uncertainty in the United States and Europe on the short-term pricing model for gold.
The study conducted by \cite{junttila2018economic} investigated the impact of EPU on predicting future real economic activities in the UK and European economies. For an extensive analysis and thorough examination of different methods for quantifying uncertainty, as well as the impact of economic policy uncertainty on various economic activities such as stock market returns, corporate capital investments and spending, corporate finance, and risk management \cite{al2019economic}. On the other hand, the effect of geopolitical risk (GPRC) has been studied extensively by \cite{caldara2022measuring}. A recent study by \cite{pringpong2023geopolitical} aims to link geopolitical risk with the reduction of firm values as well. The studies covering the impact of various uncertainty measures on different economic activities are still in the nascent phase, and this study, in turn, presents a way to accommodate the roles of EPU and GPRC in forecasting inflation numbers.

Wavelet analysis has shown remarkable promise in analyzing financial time series due to its ability to distillate crucial information from the data. Their structural flexibility allows them to handle irregular time series data more effectively for long-term time series forecasting \cite{percival1994long}. Broadly, wavelets decompose the original time series into two sub-components: one concerning the low-frequency or deterministic component and the other concerning the high-frequency or stochastic component of the series. This process aims to reduce the effect of `noise' in the forecast and is found to be quite useful in dealing with a variety of non-stationary signals, a common feature of financial and macroeconomic time series data, as they are constructed over finite time intervals and can represent the `local' elements in both time and scale \cite{percival2000wavelet}. Several applications of wavelet analysis can be found in forecasting exchange rates, GDP growth, crude oil prices, stock market prices, Taiwan stock index, S\&P dividend yield, trade prices, expenditure and income, price movements, money growth, and inflation \cite{yousefi2005wavelet, shrestha2005real}, volatility in foreign exchange markets, car sales from the last decades \cite{crowley2007guide}. A wide variety of hybrid models combining wavelets and neural networks have been studied for forecasting natural gas prices \cite{jin2015forecasting}, future trading \cite{zhang2001multiresolution}, short-term electricity load \cite{zhang2001adaptive, benaouda2006wavelet}, energy price forecasting \cite{tan2010day}, oil prices \cite{yousefi2005wavelet}, stock index \cite{dunis2002modelling}. Several studies have employed wavelet-based forecasting methods in various other domains of economics, business, and finance. For instance, \cite{fernandez2007wavelet} focused on the U.S. metal and materials manufacturing industry, employing wavelet decomposition combined with support vector machines to enhance forecasting accuracy. While the study effectively captures complex dynamics in manufacturing, it does not address long-term forecasting horizons or the inclusion of exogenous variables, which are crucial for economic forecasting. \cite{rua2017wavelet} used wavelet-based multiscale principal component analysis to improve factor model estimation and forecasting of macroeconomic variables like GDP growth and inflation in the U.S. Another paper by \cite{souropanis2023forecasting} enhanced realized volatility (RV) forecasts for the S\&P 500 using wavelet decomposition coupled with autoregressive model. However, these approaches do not address the non-linearity issue, as evident in CPI inflation data. In recent work, \cite{he2021wavelet} used a deep learning framework enhanced by wavelet transform to predict multivariate time series. Although their model captures the long-term and short-term characteristics of time series in the temporal domain through convolutional neural networks (CNN), it cannot be generalized well for limited inflation time series data problems. \cite{vogl2022forecasting} utilized the efficacy of combining wavelet functions with neural networks for financial market predictions. More recently, \cite{boubaker2023hybrid} proposed a hybrid model based on ARFIMA and wavelet neural network for forecasting the DJIA Index. However, the exclusion of exogenous variables and focus on high-frequency data limits its applicability to macroeconomic forecasting of CPI inflation. Apart from forecasting applications, wavelet analysis has been used in various areas of finance: \cite{kim2005relationship} analyzed the association between stock price and inflation rate using wavelet analysis; \cite{rua2012wavelet} explored the aspect of assessment of the risk in the emerging markets; \cite{he2012ensemble} studied the value at risk in the metal markets. \cite{alexandridis2020global} examined the impact of the global financial crisis on the multi-horizon nature of systematic risk and market risk using a wavelet multi-scale approach. \cite{becerra2005neural} used wavelet neural networks for the prediction of corporate financial distress. However, none of the above studies have utilized the inflation series and its causal counterparts for long-term forecasting.

Against this background, the primary aim of this study is to create a filtered ensemble wavelet neural network (FEWNet) model for CPI inflation forecasting. The key objective here is to generate accurate long-term forecasts for CPI inflation in BRIC countries while taking into account EPU and GPRC. FEWNet's ability to handle non-linear, non-stationary, and complex time series further underscores its robustness. Its relatively stronger performance against baselines ranging from classical time series to modern deep learning architectures establishes its superiority in inflation forecasting contexts. Additionally, macroeconomic policy variables, characterized by low-frequency data, often present challenges where methods effective for high-frequency data (particularly various neural network architectures) perform sub-optimally. % We address these shortcomings in a filtered ensemble wavelet neural network (FEWNet) model. 
In analyzing macroeconomic policy variables like CPI inflation, the wavelet decomposition can extract the low-frequency components of the series without sacrificing the fundamental characteristics of the original time series. Our model also integrates a layer for feature engineering aspects by extracting the trend components using the Hodrick-Prescott (HP) filter and cyclical elements using the Christiano-Fitzgerald (CF) ideal band-pass filter in the forecasting exercise. Initially, the observed time series (CPI inflation) and exogenous factors (EPU and GPRC) undergo decomposition into their trend and cyclical components using HP and CF filters in the first phase. Subsequently, in the second phase, all the trend and cyclical components are modeled as auxiliary inputs within the FEWNet framework, along with the wavelet-transformed details and smooth coefficients of the CPI inflation. This comprehensive approach aims to generate forecasts for 12 months and 24 months ahead. Conditioning inflation forecasts on policy uncertainty and geopolitical risk would capture the inflation or deflation effects of uncertainty. We argue using wavelet coherence analysis that there exists a relationship between EPU and GPRC with that of CPI inflation for BRIC economies. FEWNet generates out-of-sample forecasts in a recursive manner, and using several statistical metrics and tests, we compare its performance with alternative forecasting methods. Theoretical results on empirical risk minimization are also presented for the FEWNet. Our study evaluates a wide array of neural network architectures, along with several statistical benchmarks, and demonstrates that FEWNet delivers superior forecasting performance for both semi-long-term (12 months ahead) and long-term (24 months ahead) forecast horizons for CPI inflation series in the BRIC countries. Apart from improvement in forecast accuracy measures of CPI inflation for the BRIC countries, another critical area that interests the policymakers is the knowledge of how various forms of structural uncertainties can influence the movement or complex dynamic behavior of different macroeconomic variables. The performance of FEWNet is superior to its counterparts in all three economies, except for China. This demonstrates that FEWNet is highly effective in handling complicated, non-stationary time series in forecasting tasks. 

The paper is organized in the following way. Section \ref{Section_Data_and_Methodology} explains the macroeconomic series considered in this study, various data pre-processing strategies, and the causal association between the uncertainty indices and CPI inflation. Section \ref{Section_Proposed_Model} focuses on the proposed methodology with a detailed overview of its architectural design, and Section \ref{ERM_FEWNet} theoretically establishes the robustness of the method from an empirical risk minimization perspective. Section \ref{Section_Experimental_Evaluation} discusses the experimental evaluation and the process of deriving a conformal prediction interval for the FEWNet framework. We discuss the policy implications this paper leads to in Section \ref{Sec_Policy_Implications}. Finally, Section \ref{Section_Conclusion} ultimately wraps up the study by presenting a thorough analysis of the main discoveries and outlining potential future directions.

\section{Data and Preliminary Analysis}\label{Section_Data_and_Methodology}
This work is premised on forecasting the monthly CPI inflation series for the four countries: Brazil, Russia, India, and China, based on their historical data and various economic uncertainty measures such as EPU and GPRC.
% \footnote{Two news-based measures of uncertainty are used in this study: the geopolitical risk uncertainty index (GPRC) \cite{caldara2022measuring} and the economic policy uncertainty index (EPU) of \cite{baker2016measuring}. EPU stands for Economic Policy Uncertainty, which represents the likelihood of changes in monetary, fiscal, or regulatory policies that can impact the decision-making behavior of investors and consumers \cite{baker2016measuring}. The geopolitical risk index\cite{caldara2022measuring} measures the occurrence of geopolitical events such as terrorism, local and regional political instability, political violence, coups d'état, territorial conflicts, and war. These events contribute to the uncertainty in the geopolitical landscape. Please refer to \url{http://www.policyuncertainty.com/} and \url{https://www.matteoiacoviello.com/gpr.htm} for more information on the EPU and GPRC index calculations.}. 
To evaluate the performance of our proposed method, we consider two forecast horizons, namely, a semi-long-term length of 12 months and a long-term length of 24 months, for each of the countries. This section details the datasets used in our analysis, their properties, the steps involved in data preparation (Section \ref{Sec_Data_Prep}), and the causality analyses conducted (Section \ref{Causal_anal}). 
%Furthermore, the implementable codes and the datasets used in this study are made available on \href{https://github.com/ctanujit/FEWNet}{GitHub repository} for the reproducibility of the results.

\subsection{Data Preparation and Global Charateristics}\label{Sec_Data_Prep}
In this study, we use monthly data on CPI inflation, EPU, and GPRC for BRIC countries from January 2003 (2003-01) to November 2021 (2021-11). For the semi-long-term forecasting task, we set the training period from 2003-01 to 2020-11 and the test horizon as 2020-12 to 2021-11, whereas for the long-term forecasting, our training data spans from 2003-01 to 2019-11 and test period extends from 2019-12 to 2021-11. Our target CPI inflation (also headline inflation) is measured as the year-on-year percentage change in the consumer price index (CPI) series obtained from the Federal Reserve Economic Data (FRED) \url{https://fred.stlouisfed.org/}. These numbers are seasonally unadjusted and are calculated based on the following approach:
\begin{equation}
\pi_{\text {Jan }}^{2003}=\left[\frac{C P I_{J a n}^{2003}-C P I_{\text {Jan }}^{2002}}{C P I_{J a n}^{2002}}\right] * 100 \%,
\end{equation}
where $\pi_{Jan }^{2003}$ denotes the derived CPI inflation number for 2003-Jan, calculated using CPI values of 2003-Jan  and 2002-Jan (prior year and same month). 

We use two of the most popular news-based measures of uncertainty: the EPU index developed by \cite{baker2016measuring} and the GPRC index developed by \cite{caldara2022measuring}. EPU represents the likelihood of changes in monetary, fiscal, or regulatory policies that can impact the decision-making behavior of investors and consumers \cite{baker2016measuring}. According to \cite{baker2016measuring}, the EPU Index is constructed by analyzing the frequency of certain terms in newspaper articles from a variety of sources, including the ``Access World News Bank Service''. The index captures articles that mention key terms related to economic and policy uncertainty. Specifically, it searches for combinations of terms such as `economics', `deficit', `uncertainty', `monetary policy', `regulation', `legislation', `White House', `Federal Reserve', or `Congress'. This electronic text search process helps quantify the level of policy uncertainty by tracking the presence of these terms over time in major newspapers. The data on EPU is sourced from \url{http://www.policyuncertainty.com/}. Similarly, GPRC measures the occurrence of geopolitical events such as terrorism, local and regional political instability, political violence, coups, territorial conﬂicts, and war. These events contribute to the uncertainty in the geopolitical landscape. According to \cite{caldara2022measuring}, the GPRC index is constructed by conducting electronic text searches of specific terms related to geopolitical tensions across 10 prominent daily newspapers. The GPRC index categorizes search terms into eight distinct categories to measure various facets of war-related events and geopolitical tensions: War Threats (1), Peace Threats (2), Military Buildups (3), Nuclear Threats (4), Terror Threats (5), Beginning of War (6), Escalation of War (7), Terror Acts (8). The data on GPRC is obtained from \url{https://www.matteoiacoviello.com/gpr.htm}. 

As part of the data pre-processing step, we perform the log transformation on the EPU series to stabilize its variance. Sequentially, we decompose the CPI inﬂation, log-transformed EPU, and GPRC series using the Hodrick and Prescott (HP) filter \cite{hodrick1997postwar} and the Christiano-Fitzgerald (CF) random walk filter \cite{christiano2003band}. These economic filters play a vital role in extracting the trends and business cycles associated with the time series. Further details on the HP filter and CF filter are provided in Appendix \ref{App_Exo_fil}. The summary statistics of our variables under study are provided in Table \ref{Tab_General_Statistics}.

\begin{table}[h!]
\centering 
   \caption{Summary statistics of the datasets used in this study}
\footnotesize
\centering
 \begin{tabular}{ccccccccccc} \hline
Countries & Series & Obs. & Min. Value & Max. Value & ${Q}_{1}$ & Median & Mean & ${Q}_{3}$ & CoV & Entropy \\ \hline
Brazil & \textit{CPI Inflation} & 227 & 1.88 & 17.24 & 4.31 & 5.7 & 6.11 & 6.86 & 46.86 & 5.41 \\
 & \textit{log(EPU)} & 227 & 1.35 & 2.83 & 2.00 & 2.15 & 2.16 & 2.3 & 10.94 & 5.65 \\
 & \textit{GPRC} & 227 & 0.01 & 0.23 & 0.03 & 0.04 & 0.05 & 0.06 & 66.08 & 15.13 \\ \hline
 Russia & \textit{CPI Inflation} & 227 & 2.2 & 16.93 & 5.46 & 7.61 & 8.44 & 11.60 & 46.01 & 5.38 \\
 & \textit{log(EPU)} & 227 & 1.38 & 2.90 & 1.96 & 2.15 & 2.15 & 2.35 & 13.28 & 5.65 \\
 & \textit{GPRC} & 227 & 0.22 & 2.23 & 0.42 & 0.58 & 0.67 & 0.83 & 51.46 & 6.05 \\ \hline
 India & \textit{CPI Inflation} & 227 & 1.08 & 16.22 & 4.51 & 6.09 & 6.69 & 8.70 & 43.43 & 5.41 \\
 & \textit{log(EPU)} & 227 & 1.37 & 2.45 & 1.72 & 1.90 & 1.90 & 2.06 & 11.70 & 5.68 \\
 & \textit{GPRC} & 227 & 0.06 & 0.89 & 0.14 & 0.18 & 0.20 & 0.24 & 44.93 & 7.86 \\ \hline
 China & \textit{CPI Inflation} & 227 & -1.79 & 8.81 & 1.51 & 2.11 & 2.52 & 3.21 & 75.55 & 5.47 \\
 & \textit{log(EPU)} & 227 & 1.42 & 2.99 & 1.96 & 2.18 & 2.24 & 2.51 & 16.97 & 5.63 \\
 & \textit{GPRC} & 227 & 0.21 & 1.52 & 0.36 & 0.47 & 0.53 & 0.65 & 46.05 & 6.27 \\ \hline     
  \end{tabular}
   \label{Tab_General_Statistics}
\end{table}

In the analysis, we study the global features of the CPI inflation series, log-transformed EPU series, and  GPRC series to identify their structural patterns and detect appropriate techniques for long-term forecasting. Primarily, we focused on seven popular time series characteristics, including skewness, kurtosis, non-linearity, long-range dependence, seasonality, stationarity, and outlier check \cite{hyndman2018forecasting}. We utilize the Hurst exponent to assess the long-range dependency of the time series. To analyze the non-linear characteristics of the series, we employ Tsay's test and Keenan's one-degree test for non-linearity. To determine the stationarity of the series, we conduct the Kwiatkowski–Phillips–Schmidt–Shin test. Lastly, we apply Ollech and Webel's test to identify any seasonal patterns in the series.
% For evaluating the long-range dependency of the time series, we consider the Hurst exponent; the non-linearity of the series is examined using the Tsay test, for detecting the stationarity we perform the Kwiatkowski–Phillips–Schmidt–Shin test, and for the seasonality of the series we applied the Ollech and Webel's test. 
The results of the statistical features, summarized in Table \ref{Table_Global characteristics}, depict that most of the macroeconomic time series are non-stationary except for the GPRC series of Brazil and the CPI inflation data of China. The majority of these series are positively skewed except for the log-transformed EPU series of Russia. Furthermore, most of the economic time series exhibit non-linear patterns, which plays a crucial role in selecting the appropriate modeling approach for the dataset. Additionally, based on Ollech and Webel's seasonality test and the Hurst exponent we observe that all the series are non-seasonal and are characterized by long-range dependency. Moreover, the ACF and PACF plots of the CPI inflation series (presented in Table \ref{PLOTS:ACF}) depict the presence of serial autocorrelation. These inherent patterns of the CPI inflation series motivate the design and application of hybrid forecasting approaches for long-term forecasting. We also performed the Bonferroni outlier test \cite{weisberg1982residuals} based on studentized residuals. Our findings indicate that none of the CPI inflation series has significant outliers based on the adjusted p-values. However, while inflation remains relatively steady, policy uncertainty and geopolitical risks experience sudden and substantial fluctuations. This differential behavior underscores the necessity for policymakers to vigilantly monitor these indicators and implement strategic interventions aimed at mitigating the adverse effects of such outliers on the broader economy. We further examined the presence of structural breakpoints in the CPI inflation series to account for regime shifts and their potential impact on our downstream forecasting models, utilizing the ordinary least squares (OLS)-based CUSUM test \cite{ploberger1992cusum}. By examining the cumulative sum of recursive residuals, the test can reveal deviations from the model's stability, indicating potential breakpoints. In our study, the application of the OLS-based CUSUM test allows us to isolate and identify periods of structural change in the CPI inflation series. The CPI inflation series for these countries does not exhibit statistically significant breakpoints. The absence of statistically significant breakpoints in the CPI inflation series for Brazil, Russia, India, and China suggests that these series have remained relatively stable over the observed period. This stability implies that regime shifts are unlikely to pose significant challenges for forecasting models based on these series.

\begin{table*}[!ht]
\tiny
\centering
\caption{Global characteristics of the economic time series under study for BRIC countries.}
 \begin{tabular}{ccccccccc} \hline
Countries & Series & Skewness & Kurtosis & Non-Linearity & Long-Range & Seasonality & Stationarity & Outlier \\
& & & & & Dependence & & & \\ \hline
Brazil & CPI Inflation & 1.67 & 3.57 & Non-linear & 0.73 & Non-seasonal & Non-stationary  & No outliers detected\\
 & log(EPU) & -0.9 & 0.29 & Non-linear & 0.77 & Non-seasonal & Non-stationary  & 1 outlier detected\\
 & GPRC & 2.06 & 6.60 & Non-linear & 0.63 & Non-seasonal & Stationary & 4 outliers detected \\ \hline
 Russia & CPI Inflation & 0.28 & -0.98 & Linear & 0.78 & Non-seasonal & Non-stationary  & No outliers detected\\
 & log(EPU) & -0.03 & -0.12 & Non-linear & 0.79 & Non-seasonal & Non-stationary & No outliers detected \\
 & GPRC & 1.41 & 2.46 & Linear & 0.79 & Non-seasonal & Non-stationary & 2 outliers detected \\ \hline
 India & CPI Inflation & 0.71 & 0.17 & Linear & 0.82 & Non-seasonal & Non-stationary & No outliers detected\\
 & log(EPU) & 0.12 & -0.47 & Linear & 0.80 & Non-seasonal & Non-stationary & No outliers detected\\
 & GPRC & 2.51 & 15.22 & Non-linear & 0.73 & Non-seasonal & Non-stationary & 1 outlier detected \\ \hline
 China & CPI Inflation & 0.82 & 1.08 & Non-linear & 0.69 & Non Seasonal & Stationary & No outliers detected\\
 & log(EPU) & 0.23 & -0.87 & Non-linear & 0.82 & Non-seasonal & Non-stationary & No outliers detected\\
 & GPRC & 1.28 & 1.51 & Non-linear & 0.80 & Non-seasonal & Non-stationary & 1 outlier detected\\ \hline
     
  \end{tabular}
  \label{Table_Global characteristics} 
      % \end{tcolorbox}
\end{table*}

\begin{table*}[!ht]
    \small 
    \centering
    \caption{Training data for FEWNet including target series CPI inflation (blue), and exogenous variables log-transformed EPU (red) and GPRC (green). ACF, PACF, and OLS-based CUSUM test plots of CPI inflation series of the BRIC countries.}
    \begin{tabular}{ p{5.8cm}  p{3.2cm}  p{3.2cm} p{3.2cm} }
        \hline
        % Country & 
        Training data (Jan 2003 to Nov 2019) & ACF plot & PACF plot & OLS-based CUSUM test  \\ \hline
        %         &
        \multicolumn{4}{c}{\begin{minipage}{\textwidth}
            \includegraphics[width=168mm, height=29mm]{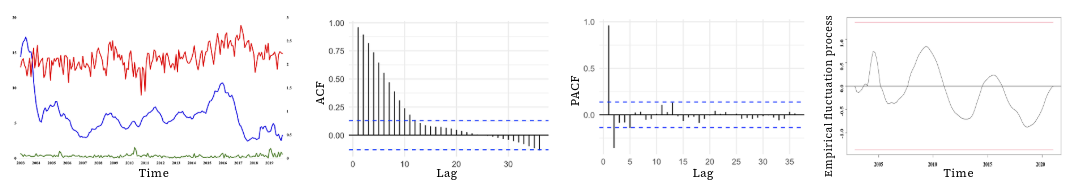}
        \end{minipage}}\\ 
        \multicolumn{4}{c}{Brazil} \\ \hline
        \multicolumn{4}{c}{\begin{minipage}{\textwidth}
            \includegraphics[width=168mm, height=29mm]{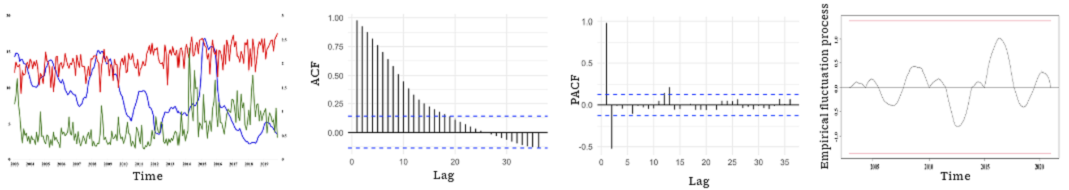}
        \end{minipage}}\\ 
        \multicolumn{4}{c}{Russia} \\ \hline
        \multicolumn{4}{c}{\begin{minipage}{\textwidth}
            \includegraphics[width=168mm, height=29mm]{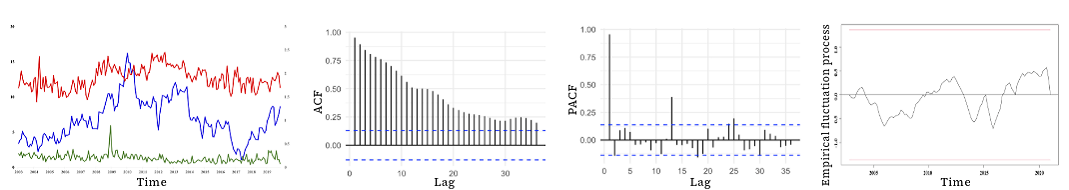}
        \end{minipage}}\\ 
        \multicolumn{4}{c}{India}\\ \hline
        \multicolumn{4}{c}{\begin{minipage}{\textwidth}
            \includegraphics[width=168mm, height=29mm]{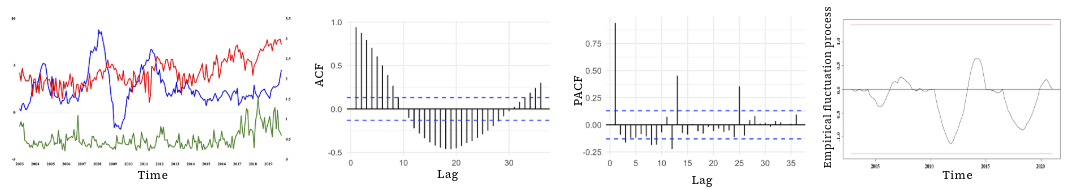}
        \end{minipage}}\\ 
        \multicolumn{4}{c}{China}\\ 
        \multicolumn{4}{c}{\begin{minipage}{\textwidth}
            \includegraphics[width=60mm, height=4mm]{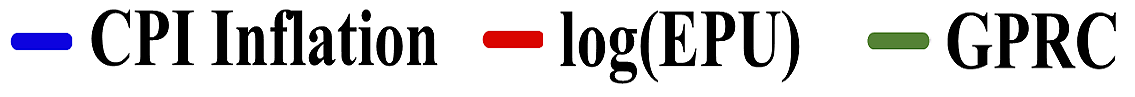}
        \end{minipage}}\\ \hline
    \end{tabular} 
    \label{PLOTS:ACF}
\end{table*}

\subsection{Causality Analysis}\label{Causal_anal}
This section explores the causal impact of log-transformed EPU and GPRC uncertainties on the CPI inflation series for the BRIC countries. Several statistical methods, such as the Granger causality test, cross-convergent mappings, and transfer entropy, have been suggested in the literature to examine the causal relationship between two variables \cite{pierce1977causality}. Nevertheless, the tests rely on certain data-level assumptions that are not met in the macroeconomic series analyzed in this work. In order to address this constraint, we employ wavelet coherence analysis (WCA) to examine the causal relationship between the series \cite{grinsted2004application}. The WCA methodology offers a valuable method for examining the interconnection and simultaneous movement of two non-stationary signals in the time-frequency domain.
\begin{figure}[!ht]
 \centering
  \includegraphics[width=0.9\textwidth]{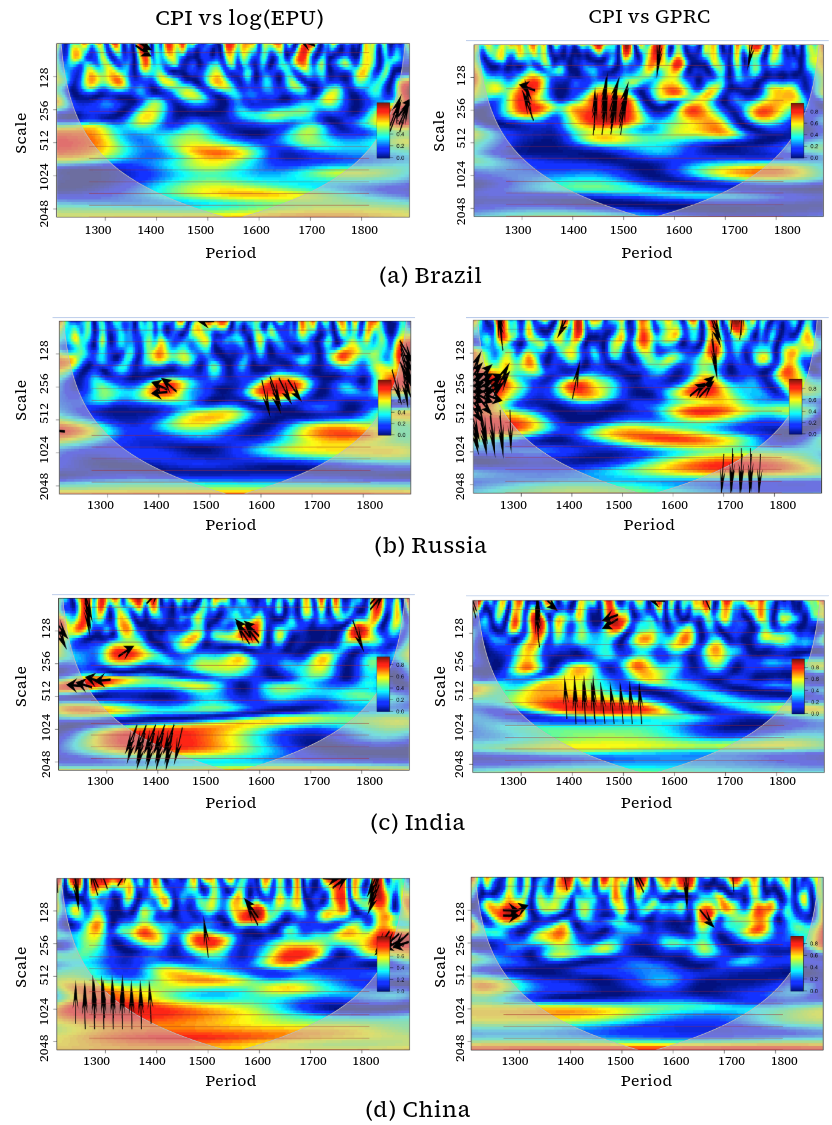}
   \caption{Wavelet coherence analysis plots of CPI inflation with log-transformed EPU (left) and CPI inflation with GPRC (right) for (a) Brazil, (b) Russia, (c) India, and (d) China.}
   \label{fig:WCA}
\end{figure}

In the WCA plots, demonstrated in Figure \ref{fig:WCA}, the first row depicts the WCA results between CPI inflation and log-transformed EPU series, and the second row shows the same for CPI inflation and GPRC series. In each of the plots, the horizontal axis measures the time, and the vertical axis shows the frequency dimension. Wavelet coherence represents the regions in the time-frequency domain where the two signals co-vary. In the plot, the transition from warmer colors (red) to colder colors (blue) indicates regions from higher coherence to lower coherence between the series. The arrow in the WCA plots indicates the lag phase relations between the analyzed signals. A zero-phase difference would hint toward the co-movement of the signals on any particular scale. The direction of an arrow has a specific connotation, with the arrows pointing to the right (left) indicating that the signals are in phase (anti-phase), i.e., the two signals are moving in the same direction or vice versa. As evident from the plots, there exists a significant causal relation between CPI inflation, log-transformed EPU, and GPRC series. This in turn, justifies the choice of EPU and GPRC as exogenous factors in forecasting the CPI inflation series of BRIC countries.

\section{Proposed FEWNet Model}\label{Section_Proposed_Model}
\begin{figure}[h!]
 \centering
  \includegraphics[width=1.00\textwidth]{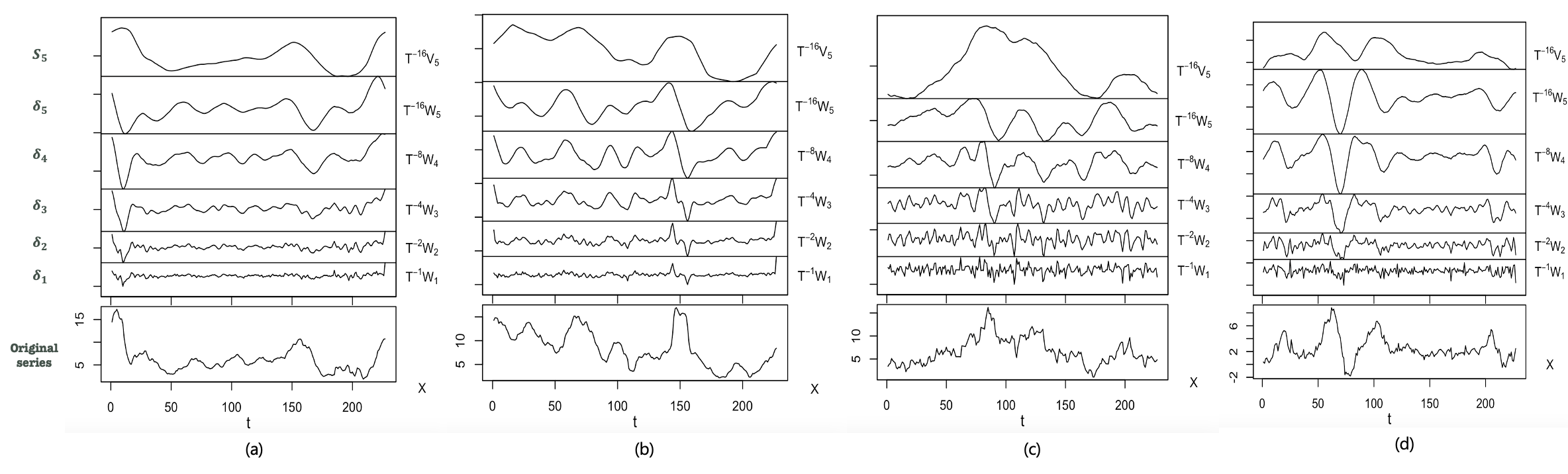}
   \caption{MODWT decomposition of the CPI Inflation series for (a) Brazil, (b) Russia, (c) India, and (d) China between the period Jan-2003 and Nov-2021. In these figures, $\delta_{1}, \delta_{2}, \ldots, \delta_{5}$ and $S_{5}$ represent the details and smooth of the series generated by the MODWT-based MRA approach. The bottom chart in all the figures represents the original time series (CPI inflation) in actual frequency scale} 
   \label{fig:MODWT_Trans}
\end{figure}
This section provides a comprehensive description of our proposed architecture, called the filtered ensemble wavelet neural network (FEWNet). FEWNet combines the maximal overlapping discrete wavelet transformation (MODWT) technique (explained in Appendix \ref{App_DWT}), economic filtering methods such as the Hodrick-Prescott (HP) filter and Christiano-Fitzgerald (CF) filter (explained in Appendix \ref{App_Exo_fil}), and the autoregressive neural network with exogenous variables (ARNNx). The FEWNet framework is a sequential approach that initially decomposes the CPI inflation data into low-frequency (smooth) and $\mathcal{K}$ high-frequency (detail) components using the scaling and wavelet filters of MODWT, as depicted in Figure \ref{fig:MODWT_Trans}. The ability of the transition and time-invariant MODWT approach to handle the non-stationary features of the time-dependent signal makes it a valuable tool in pre-processing the CPI inflation dataset. Simultaneously, to capture the economic uncertainties of the time series, we apply the CF and HP filters on the CPI, log-transformed EPU, and GPRC series. The non-parametric CF filter efficiently captures the trend components, whereas the HP filter is suitable for identifying and modeling the business cycles. This long-term trend and the repetitive cyclical patterns have been found to provide significant information regarding the economic uncertainties that impact the future trajectories of the CPI inflation series.

The real-world CPI inflation series exhibit notable features, including non-stationarity, non-linearity, and long-term dependence, as evidenced in Table \ref{Table_Global characteristics}. In order to accurately represent this non-uniform time series, we utilize MODWT decomposition using Daubechies orthogonal wavelets of length 2, namely the Haar filter. This transformation has the ability to identify the specific time and frequency components of signals, successfully distinguishing between long-range connections, non-stationarity, and patterns within the signals \cite{percival1997analysis}. By utilizing the MODWT-based multi-resolution analysis (MRA) decomposition on the series $Y_t$, we derive a collection of $\mathcal{K}+1$ distinct random variables called wavelet and scaling coefficients, which are not correlated with each other. In our suggested methodology, we employ separate neural networks within an ensemble framework to represent $\mathcal{K}+1$ random variables instead of employing the original CPI inflation series $Y_t$. These neural networks are specifically engineered and trained to effectively manage the intricacies linked to the wavelet coefficients. By employing this approach, we may efficiently tackle the non-linear nature observed in the CPI inflation data of the BRIC nations. In addition, in order to capture the temporal causal association between the input series and the exogenous variables ($X_{j,t}, \; j = 1, 2, \ldots, 6$), we include them in each of the $\mathcal{K}+1$ wavelet transformed series of $Y_t$ in their corresponding neural networks. The forecasts produced by these $\mathcal{K}+1$ neural networks are combined to determine the future dynamics of $Y_t$. The mathematical expression for the $m$-step ahead forecasts of $Y_t$, designated as $\hat{Y}_{\mathcal{N}+m}$, based on $\mathcal{N}$ past data, is as follows:
\begin{equation}\label{Eq_ERM_3}
    \hat{Y}_{\mathcal{N}+m} = \hat{S}_{\mathcal{K},\mathcal{N}+m} + \sum_{\tilde{k} = 1}^{\mathcal{K}} \hat{\delta}_{\tilde{k},\mathcal{N}+m} ,
\end{equation}
where $\hat{S}_{\mathcal{K},\mathcal{N}+m}$ and $\hat{\delta}_{\tilde{k},\mathcal{N}+m}, \; \tilde{k} = 1, 2, \ldots, \mathcal{K}$ are the forecasts generated from the simultaneously computed neural networks of the smooth series $S_{\mathcal{K},t}$ and the details series $\delta_{\tilde{k},t} \; \tilde{k} = 1, 2, \ldots, \mathcal{K}$. 
Each of these $\mathcal{K}+1$ neural networks consists of a feedforward design with three layers - the input layer with $p$ nodes, one hidden layer with $q$ nodes, and one output layer without any shortcut connections. The neural networks model incorporates $p-6$ lagged observations of the wavelet coefficients, as well as one lagged observation for each of the six exogenous variables ($X_j$). These variables represent the trend and cycle information of CPI, log-transformed EPU, and GPRC. This setting allows the networks to generate one-step forward forecasts for the related series. Mathematically, the result obtained from each neural network after one iteration can be represented as:
\begin{equation}\label{Eq_2}
 \begin{gathered}
    \hat{\delta}_{\tilde{k}, \mathcal{N}+1} = \alpha_{0, \tilde{k}} + \sum_{i = 1}^q \beta_{i, \tilde{k}} \sigma(\alpha_{i, \tilde{k}} + \beta^{'}_{i, \tilde{k}} \vec{\delta}_{\tilde{k}}  + \gamma^{'}_{i, \tilde{k}} \underbar{X}); \; \tilde{k} = 1, 2, \ldots, \mathcal{K} \\
    \hat{S}_{\mathcal{K}, \mathcal{N}+1} = \alpha_{0}^{S} + \sum_{i = 1}^q \beta_{i}^{S} \sigma(\alpha_{i}^S + {\beta^S_{i}}^{'} \vec{S}_{\mathcal{K}} + {\gamma^{S}_{i}}^{'} \underbar{X}),
\end{gathered}    
\end{equation}
where $\vec{\delta}_{\tilde{k}}$ and $\vec{S}_{\mathcal{K}}$ denotes the $p-6$ lagged observation of the corresponding wavelet coefficients and $\underbar{X}$ indicates a single lagged vector of exogenous variables, $\alpha_{0, \tilde{k}}, \alpha_{0}^{S}, \alpha_{i, \tilde{k}}, \alpha_{i}^S$ are the bias terms, $\beta_{i, \tilde{k}}, \beta_{i}^{S}$ are the connection weights between hidden and output layer, $\gamma^{'}_{i, \tilde{k}}, {\gamma^{S}_{i}}^{'}$ are the weight vectors between the exogenous variables and the hidden layer, $\beta^{'}_{i, \tilde{k}}, {\beta^S_{i}}^{'}$ are the connection weights between lagged wavelet series and the hidden layer, and $\sigma$ indicates the non-linear activation function. 
Within our framework, we commence by assigning random initial values to the connection weights. Subsequently, we employ the gradient descent back-propagation approach to train these weights, as described in the work by \cite{rumelhart1986learning}. The aforementioned technique produces a forecast that is one step ahead for each of the wavelets and scaling coefficients. In order to make forecasts for several future time steps, we incorporate the most recent predictions into the input layer instead of using the last observed value, and we continue this procedure recursively. Ultimately, we combine the predictions produced by each of the base models in an ensemble setup and derive the forecast for a for a desired time horizon.

\begin{figure}[h!]
 \centering
  \includegraphics[width=\textwidth]{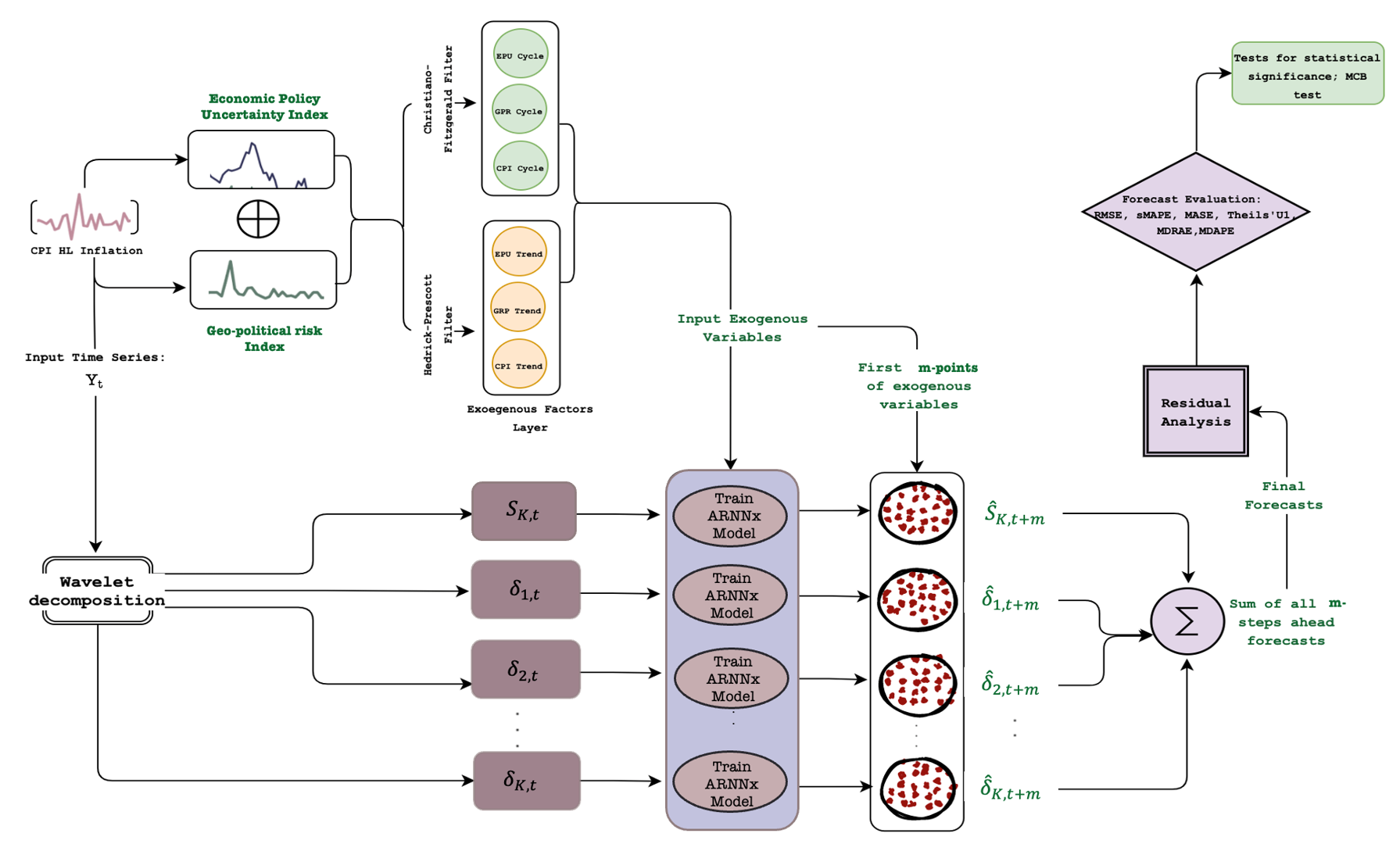}
   \caption{Illustration of the proposed FEWNet model} 
   \label{fig:table_8}
\end{figure}

The FEWNet algorithm consists of three hyperparameters: the level of MODWT decomposition ($\mathcal{K}$), the number of lagged observations in the input layer ($p$), and the number of hidden nodes ($q$). Prior research has indicated that selecting the appropriate level of wavelet decomposition ($\mathcal{K}$) is vital for distinguishing the genuine signal from noise in a complicated series \cite{percival1997analysis, aussem1997combining}.
% Thus to restrict the computational complexity of the shift-invariant MODWT approach we set 
In order to limit the computational cost of the shift-invariant MODWT technique, we choose $\mathcal{K} = \lfloor\log_e\text{(length of training set}) \rfloor = \lfloor\log_e \mathcal{N} \rfloor$, where $\lfloor \cdot \rfloor$ denotes the floor function, following \cite{percival2000wavelet}. To specify the number of previous lagged observations in the input layer, we utilize the cross-validation strategy and select $p$ so that it minimizes the symmetric mean absolute percent error (SMAPE) of the validation set, i.e.,
\begin{equation}
    p = \underset{p}{argmin} \frac{1}{|\mathcal{V}|}\sum_{t \in \mathcal{V}}\frac{2|\hat{Y}_t - Y_t|}{|\hat{Y}_t| + |Y_t|},
\end{equation}
where $\mathcal{V}$ indicates the validation set and $Y_t$ and $\hat{Y}_t$ are the original series and forecast at time $t$. Moreover, to stabilize the learning rate of the neural network, restrict overfitting, and to reduce the run-time of the proposed FEWNet framework, we set the number of hidden nodes arranged in a single hidden layer as $q = \left[\frac{p+1}{2}\right]$ following \cite{hyndman2018forecasting, panja2023epicasting}. 
% A schematic workflow of the proposed FEWNet model is presented in Fig. \ref{fig:table_8}. As the figure depicts, the target time series is first transformed using a MODWT-based MRA into several levels. 
The suggested FEWNet model is illustrated in Figure \ref{fig:table_8} using a schematic approach. The target time series is initially subjected to a MODWT-based MRA, resulting in multiple degrees of transformation, as shown in the figure.
The economic uncertainties and geopolitical risk index series are filtered along with the CPI inflation series using the CF filter and HP filter to generate the corresponding cycle (green circles) and trend (yellow circles), respectively. These six exogenous variables are then provided with the lagged wavelet series in the input layer of each of the $\mathcal{K}+1$ neural networks. 
% The connection weights connect each input set to the hidden nodes (line with arrows). The outputs from the hidden stack are processed in the output layer, which eventually generates the one-step ahead forecast from the interconnected network.
The connection weights establish the connections between each input set and the hidden nodes, represented as lines with arrows. The results from the hidden layer are processed in the output layer, ultimately producing the forecast for the next step from the interconnected network.
% Finally, multi-step ahead forecasts are generated in a recursive manner and ensembled to produce the future dynamics of the CPI inflation series. Applying wavelet decomposition to reveal the complex variability of the transformed series and modeling them with various economic filtered uncertainties using an auto-regressive neural network with auxiliary information in an ensemble framework justifies the nomenclature of the proposed model as FEWNet. The implementation technique of the proposed FEWNet is summarized in Algorithm \ref{algo}.
Ultimately, the future dynamics of the CPI inflation series are formed by producing multi-step ahead forecasts in a recursive fashion and then combining them in an ensemble. The utilization of wavelet decomposition to reveal the complex variability of the transformed series and the subsequent modeling of these fluctuations with diverse economic filtered uncertainties using an auto-regressive neural network with auxiliary information within an ensemble framework provides a valid rationale for naming the proposed model as FEWNet. The implementation methodology of the proposed FEWNet is outlined in Algorithm \ref{algo}.

%An overview and detailed description of the complete implementation steps of the proposed model (FEWNet) is available in \boldsymbol{Algorithm 1}:
\begin{remark}
    Conventional machine learning and deep learning frameworks, such as XGBoost and TFT, often use a sliding window approach to transform time series forecasting into a supervised learning task; our proposed FEWNet architecture takes a distinct approach. We utilize an ensemble ARNNx framework on the MODWT decomposed training series and economically filtered exogenous variables. Unlike typical methods, the ARNNx model doesn't reconstruct the economic series into an input-output supervised framework; instead, it leverages p-6 lagged observations (as we have 6 exogenous variables after HP and CF decomposition) of each of the decomposed training data and the previous lagged values of the exogenous variable to generate a one-step-ahead forecast using a non-linear function as given in Eq. (\ref{Eq_ERM_3}) and (\ref{Eq_2}). Furthermore, we iteratively update the training data with the latest forecast, developing a multi-step ahead forecast for each transformed series and aggregating forecasts from each wavelet decomposed series for our results. Our experimental setup involves two test horizons (24 months and 12 months) and distinct training periods corresponding to each of the horizons. For the 24 months period, we consider the training data from January 2003 – November 2019 and perform the decompositions of this available training dataset. For the 12-month horizon, the training period spans from January 2003 to November 2020, and the proposed FEWNet framework is fitted into this training set. Finally, for evaluation purposes, we use the original test data from each of these forecast horizons only to report the efficiency of the proposed FEWNet framework. Therefore, there is no data leakage while decomposing the data in this paper. 
\end{remark}

\begin{algorithm}
%Offline phase: Train the learning models (FFNN, RC and LSTM).\\
\KwIn{The training data for the FEWNet model including the target time series $\{Y_t, \; t = 1, 2, \ldots \mathcal{N}\}$ indicating CPI inflation and exogenous series $X_{t}$ comprising of log-transformed EPU, and GPRC.} 
\KwOut{$\hat{Y}_{\mathcal{N}+1}, \hat{Y}_{\mathcal{N}+2}, \ldots, \hat{Y}_{\mathcal{N}+m}; \; m \; (\geq 1)$ indicating $m$ step-ahead forecast of the CPI inflation series.}
\vspace{0.2cm}
{\textbf{Methodology:}}\\
\nl Transform the target data vector ($Y_t$) into $\mathcal{K}$ levels by applying MODWT with haar filter to generate a sequence of $\mathcal{K}+1$ uncorrelated random variables indicating the local trend and high-frequency fluctuations of the series. \\
% \nl Select the value of $\mathcal{K}$ as $\operatorname{log}_e(\mathcal{N})$, where $\mathcal{N}$ indicates the size of the training data. \\
\nl Determine the value of $\mathcal{K}$ as $\lfloor\operatorname{log}_e(\mathcal{N})\rfloor$, where $\mathcal{N}$ represents the size of the training dataset. \\
\nl To quantify the economic uncertainty parameters of inflation data, apply CF and HP filters on the exogenous variables and the target variable to generate their corresponding cyclical and trend behavior, respectively. \\
\For{$\tilde{k} \leftarrow 1$ \KwTo $\mathcal{K}$}{
       % Model $\tilde{k}^{th}$ detail series with $X_{j, t},\; j = 1, 2, \ldots, 6$ using an auto-regressive neural network. \\
       Utilize an auto-regressive neural network to model the $\tilde{k}^{th}$ detail series with $X_{j, t},\; j = 1, 2, \ldots, 6$. \\
       % Design the feed-forward neural network with one input layer having $p$ nodes, one hidden layer with $q$ neurons, and an output layer. \\
       Build a feed-forward neural network consisting of an input layer with $p$ nodes, a hidden layer with $q$ neurons, and an output layer. \\
       % Input one-lagged observation for each of the six filtered covariates and $p-6$ lagged values of the $\tilde{k}^{th}$ detail series in the neural network, which subsequently passes through $q$ hidden neurons and generates a one-step-ahead forecast.  \\
       Input one-lagged observation for each of the six filtered covariates along with $p-6$ lagged values of the $\tilde{k}^{th}$ detail series as input to the neural network. The neural network will then pass this input via $q$ hidden neurons and generate the forecast for the next time step. \\
       % Select $p$ by minimizing the SMAPE and set $q =  \left [\frac{\displaystyle{p+1 }}{\displaystyle{2}}\right ]$, for stable learning architecture and shorter run time of the model. \\
       Determine the value of $p$ that minimizes the SMAPE and set $q =  \left [\frac{\displaystyle{p+1 }}{\displaystyle{2}}\right ]$, for stable learning architecture and shorter run time of the model. \\
       \KwRet{One-step forward forecast of the details series.} \\
    } %with 'haar' filter, number of levels being set to Waveletlevels and 'periodic' boundary condition. 
%Following the transformation, we 
% \nl Model the smooth series of $Y_t$ and the exogenous variables with a similar auto-regressive neural network as defined for the details series and generate the subsequent forecast. \\
\nl Apply an auto-regressive neural network to both the smooth series of $Y_t$ and the exogenous variables, following the same definition used for the details series. Use this model to generate a forecast for future values. \\
% \nl Form an ensemble framework to combine the forecasts generated from different series and formulate the final one-step-ahead prediction. \\
\nl Create an ensemble framework to merge the forecasts produced from several series and get the final one-step forward prediction. \\
% \nl Generate multi-step ahead forecast iteratively. \\
\nl Iteratively generate multi-step forward forecasts. \\
\vspace{0.2cm}
%\vspace{0.2cm}
\caption{{\bf Proposed FEWNet model}}
\label{algo}
\end{algorithm}

\section{Empirical Risk Minimization}\label{ERM_FEWNet}
FEWNet utilizes wavelet decomposition as a filtering stage and combines it with autoregressive neural networks that incorporate exogenous variables. This combination allows for precise predictions of time series that are both non-stationary and non-linear. Wavelet decomposition yields a hierarchical structure of new time series derived from the original time series, facilitating their modeling and forecasting. This section demonstrates that FEWNet effectively minimizes the empirical risk compared to classical methods, hence highlighting its practical relevance.

Let $Y_t, \; (t = 1, 2, \ldots, \mathcal{N})$ be the observed time series (stationary if random) and we aim to forecast the values of $Y_{p+1}, Y_{p+2}, \ldots, Y_{p+m}$ ($m \geq 1$ is the forecast step) using $p$ observations iteratively. In regression setup, a functional relationship is established between $Y_{p+m}$ and the input vector 
[$Y_1, Y_2, \ldots, Y_p$] by minimizing the predictive error. The optimal prediction sequence $\hat{Y}_{p+1}, \hat{Y}_{p+2}, \ldots$ minimizes the generalization risk defined as follows
\begin{equation}
    \mathcal{R}_{\text{Gen}} =  \lim_{T\to\infty} \frac{1}{T} \sum_{p = 1} ^ T \mathbb{E} \left[ \left(\hat{Y}_{p+m} - Y_{p+m}\right)^2 | Y_p, Y_{p-1}, \ldots \right], 
\end{equation}
where $\hat{Y}_{p+m} = \mathbb{E} \left[ Y_{p+m}| Y_p, Y_{p-1}, \ldots \right]$. However, computation of $\hat{Y}_{p+m}$ is difficult as the observed time series has $\mathcal{N}$ observations (training samples). Therefore, we minimize the following empirical risk: 
\begin{equation}\label{New_eq_111}
    \mathcal{R}_{\text{Emp}} =  \frac{1}{\mathcal{N}} \sum_{i = 1} ^ {\mathcal{N}} \left(Y_i - \hat{Y}_i \right)^2,
\end{equation}
where $\hat{Y}_i$ is the prediction (expectations can be omitted for deterministic time series). In practical problems of economic time series forecasting, the relationship between $\hat{Y}_{p+m}$ and the observed time sequence $Y_p, Y_{p-1}, \ldots$ is supposed to be non-linear in nature. Thus, the inclusion of autoregressive neural networks (a set of feed-forward artificial neural networks designed specifically for predicting future values in time series data) for forecasting is justified in this exercise. The relationship can be estimated by 
\begin{equation}\label{Eq_new_star}
    \hat{Y}_{p+m} = f(Y_p, Y_{p-1}, \ldots, Y_1),
\end{equation} 
where $f$ denotes an autoregressive neural network function. The model's order ($p$) is an essential parameter to estimate. Smaller values of $p$ make the model training simple but miss the information about the past lagged values, whereas larger values of $p$ compel the model training hard, resulting in a ``curse of dimensionality''. To avoid the issues of underfitting and overfitting, we use the ARNNx model, where we choose $p$ by minimizing the SMAPE for the validation set. If $f$ in Eq. (\ref{Eq_new_star}) is a complex estimator (e.g., deep learning models such as transformers, deep MLP, etc.), it will learn well on the training data but perform poorly on unseen test data. Application of MODWT decomposes the time series using low and high pass band filters iteratively, resulting in a trend series and several detailed series (containing dynamics of the economic system at different scales). ARNNx is then trained to model the trend and detailed series and provide predictions for future values. The combined values of these MODWT-decomposed series provide a prediction of the original economic data. Although the main motivation for using wavelet decomposition on these economic time series is to enhance the predictive power, MODWT gives the smooth term $S_{\mathcal{K}}$ that contains the slowest dynamics (noise-free) and $\delta_{\tilde{k}}$ that contains system dynamics (noisy). 
The low-resolution data are frequently contaminated by noise. Consequently, training an ARNNx model on these time series is less challenging compared to the original time series. As $\tilde{k}$ increases, the dynamics get slower, resulting in low-level detail series that consist solely of noise. To mitigate overfitting, one can assign a value of zero to their predictions.
% The low-detail series are often corrupted by noise. Due to this, training an ARNNx model on these time series is easier than on the original time series. For the greater value of $\tilde{k}$, the slower are the dynamics, therefore, low-level detail series contains pure noise. To avoid overfitting, one can put their predictions as zero. 

Several recent works~\cite{aussem1997combining, soltani2002use, aminghafari2007forecasting, panja2023epicasting} have proposed to treat wavelet-decomposed time series independently. Another complicated approach will be to treat each series along with other series (treated as exogenous variables) but it results in increasing the dimensionality of the time series forecasting problem. In the current formulation we model $S_{\mathcal{K}}$ and other $\delta_{\tilde{k}}, \; (\tilde{k} = 1, 2, \ldots, \mathcal{K})$ series using ARNNx models as follows:
\begin{equation}\label{Eq_New_2}
    \hat{S}_{\mathcal{K}, p+m} = \hat{f}_0 \left(S_{\mathcal{K},p}, S_{\mathcal{K},p-1}, \ldots, S_{\mathcal{K},p-r_0}, \underbar{X} \right) \text{ and } 
    \hat{\delta}_{\tilde{k}, p+m} = \hat{f}_{\tilde{k}} \left(\delta_{\tilde{k},p}, \delta_{\tilde{k},p-1}, \ldots, \delta_{\tilde{k},p-r_{\tilde{k}}}, \underbar{X} \right), \; \tilde{k} = 1, 2, \ldots, \mathcal{K}.
\end{equation}
The choice of $\hat{f}_{i} (i = 0, 1, 2, \ldots, \mathcal{K})$ is related to the properties of the time series (non-linear for this case). \cite{panja2023epicasting} proposed non-linear ARNNx model, and each  $\hat{f}_i$ may have the same order for $r_i$ (we will use $r$ for the rest) to decrease the model's complexity. The predictions are made as an equal-weighted combination of smooth and details predictions:
%$$\hat{Y}_{\mathcal{N}+m} = \hat{S}_{\mathcal{K},\mathcal{N}+m} + \sum_{\tilde{k} = 1}^{\mathcal{K}} \hat{\delta}_{\tilde{k},\mathcal{N}+m} , $$
\begin{equation}
    \hat{Y}_{p+m} = \hat{S}_{\mathcal{K},p+m} + \sum_{\tilde{k} = 1}^{\mathcal{K}} \hat{\delta}_{\tilde{k},p+m}.
\end{equation}
In this context, we establish empirical risk minimization (ERM) property for our proposed FEWNet framework. In statistical learning theory, ERM defines a family of learning models and provides theoretical bounds on their performances. This is useful since in practice we can't generalize how well a model will work (called true risk) due to not knowing the true data distribution. However, we can instead measure its performance on a known training data set (called empirical risk).

\noindent In FEWNet, we use ARNNx estimators $\hat{f}_0, \hat{f}_1, \ldots, \hat{f}_{\mathcal{K}}$ with same order $r$ having equal number of neurons in the network architecture. These estimators are obtained by simultaneously minimizing the following empirical risk:
\begin{equation}\label{EMP_4}
    \mathcal{R}_{\text{Emp}}^W = \frac{1}{\mathcal{N}} \sum_{p = 1}^{\mathcal{N}} \left[\left(S_{\mathcal{K},p+m} - \hat{S}_{\mathcal{K},p+m} \right) + \sum_{\tilde{k} = 1}^{\mathcal{K}} \left(\delta_{\tilde{k},p+m} - \hat{\delta}_{\tilde{k},p+m} \right) \right]^2.
\end{equation}
% Since ARNNx estimators are written as a non-linear combination of linear projection of lagged inputs for the CPI inflation series, the following result holds:
Since ARNNx estimators can be expressed as a non-linear combination of linear projections of lagged inputs for the CPI inflation series, the following result holds:

\begin{prop}\label{prop_1}
    Let the autoregressive neural network (of order $r$) is applied to the original data (as in Eq. (\ref{Eq_new_star})) by minimizing the risk $\mathcal{R}_{\text{Emp}}$ (as given in Eq. (\ref{New_eq_111})) and FEWNet fits the ensemble model on the decomposed data by minimizing the empirical risk $\mathcal{R}_{\text{Emp}}^W$ (as in Eq. (\ref{EMP_4})), then we have
    \begin{equation}
    \operatorname{min} \mathcal{R}_{\text{Emp}}^W \leq \operatorname{min} \mathcal{R}_{\text{Emp}}.
    \end{equation}
\end{prop}

%\begin{proof}
\noindent\textbf{Proof.} See Appendix \ref{Prop_proof} for the detailed proof.
%\end{proof}
\begin{remark}
    However, it by no means guarantees the generalization risk reduction. In practical scenarios, the selection of the models for $\delta_{\tilde{k}, p}$ and $S_{\mathcal{K}, p}$ would be crucial and can be done using cross-validation. Also the choice of $r_{i}, \; (i = 0, 1, \ldots, \mathcal{K})$ may change the results in practice as can be thought of as a future advancement of this study. Our theoretical results show the robustness of the wavelet decomposed approaches in our proposal from an ERM perspective.
\end{remark}

\section{Experimental Evaluation and Analysis of Results}\label{Section_Experimental_Evaluation}
We assessed the efficacy of the proposed FEWNet framework by comparing its performance to other baseline predictions derived from statistical, machine learning, and deep learning methodologies. In our experimental evaluation, we utilize a 5-fold time series cross-validation approach to train all the forecasting frameworks. We then generate forecasts for two different time horizons, specifically 12 months and 24 months, to demonstrate the generalizability of the setup. In this section, we briefly explain the baseline models (Section \ref{Sec_Baseline}), evaluation metrics (Section \ref{Sec_evaluation_Metric}), benchmark comparison and experimental results (Section \ref{Sec_setup_result}), the statistical significance of the results (Section \ref{Sec_stat_signif}), empirical validation of theoretical results (Section \ref{Sec_empirical_validation_of_results}), robustness and sensitivity analysis (Section \ref{sec_robustness_sensitivity}), and conformal prediction (Section \ref{Sec_conf_pred}).

\subsection{Baseline Models}\label{Sec_Baseline}
We compared the FEWNet model and various baseline models, as well as a range of machine learning and deep learning algorithms that allow for the incorporation of exogenous factors. The evaluation incorporates the following competitive models:

\begin{itemize}
\item \textit{Random Walk} (RW) is one of the simplest stochastic models based on the assumption that in each period, the time-dependent variable takes a random step away from its previous value, and the steps are independently and identically distributed in size with zero-mean \cite{pearson1905problem}. The model assumes no mean reversion, which can serve as a naive benchmark. \textit{Random Walk with Drift} (RWD) is similar to RW but includes a constant drift term.
\item \textit{Autoregressive} (AR) model forecasts the variable of interest using a linear combination of the past values of the variable. It is like a multiple linear regression with $p$ lagged values of $Y_t$ as predictors, thus referred to as AR($p$) model (autoregressive model of order $p$) \cite{hyndman2018forecasting}.
\item \textit{Markov switching generalized autoregressive conditional heteroscedasticity} (MSGARCH) model combines the generalized autoregressive conditional heteroskedasticity (GARCH) model with a Markov-switching process \cite{gray1996modeling, haas2004new}. This allows for changes in the volatility regime according to a Markov process. MSGARCH accounts for regime changes in volatility, providing a robust benchmark for comparing models in financial econometrics.

\item \textit{Autoregressive Integrated Moving Average with exogenous variable} (ARIMAx) model is widely used in time series forecasting \cite{box1970distribution}. The ARIMAx $(p,d,q)$ framework captures the linear patterns of the time series by considering $p$ previous values of the target series, $q$ prior forecast errors, and historical values of the exogenous variable. This framework employs a differencing technique of order $d$ to guarantee the stationarity of the data. The coefficients of the linear ARIMAx model are estimated by minimizing the Akaike information criterion (AIC) value.
\item \textit{Seasonal ARIMAx} (SARIMAx) is an extension of the ARIMAx model for seasonal time series datasets. The model has a similar architecture as the ARIMAx model with additional parameters for modeling the linear trend in the seasonal components of the series. Thus the SARIMAx $(p, d, q)(P, D, Q, s)$ framework models the non-seasonal components of time series with parameters $(p, d, q)$ and for the seasonal components of length $s$ it utilizes the parameters $(P, D, Q, s)$.
\item \textit{Auto-regressive Fractionally Integrated Moving Average with exogenous variable} (ARFIMAx) generalizes the classical ARIMAx model by allowing the value of the differencing parameter $d$ to be a fraction i.e. $d \in (-0.5, 0.5)$. ARFIMAx (p, d, q) process has been widely studied to model and forecast time series exhibiting long-range dependence. As evident in Table \ref{Table_Global characteristics}, the CPI Inflation series along with the economic uncertainty indices exhibit long-range dependency, thus the choice of using the ARFIMAx model seems reasonable for our analysis to account for slowly decaying auto-correlations. 
\item \textit{Deep-learning based Auto-regressive} (DeepAR) model is an advanced deep learning system designed specifically for forecasting time series data. This framework employs a recurrent neural network architecture to forecast future trajectories in a dataset that varies over time \cite{salinas2020deepar}.
\item \textit{Neural Basis Expansion Analysis for Time Series} (NBeats) is a dense neural network structure explicitly developed for time series prediction problems. This model consists of many blocks, each composed of two main layers. The initial layer is responsible for modeling the time series to replicate previous observations and produce forecasts. Conversely, the second layer focuses on re-modeling the discrepancies between the actual and predicted values generated by the first layer and adjusting the forecasted figures. In our testing, we set the number of blocks as 2 to decrease the computational complexity of the framework.
\item \textit{Auto-regressive Distributed Lag} The autoregressive distributed lag (ARDL) model has been extensively employed in the field of econometrics to estimate the long-run and short-run dynamics of time series data \cite{pesaran1995autoregressive}. The distributed lag components of the ARDL model are capable of simulating both the stationary and non-stationary time series components.
\item \textit{Lasso Regression} is a regularized linear regression model that handles overfitting using the L1 penalty. The penalty term involved in the loss function of the LASSO model helps in shrinking the coefficients of the regressor variables with limited predictive power, thus leading to feature selection. 
% The Lasso Regression method gets combined with a multi-step forecasting layer to generate the long-term forecasts for CPI inflation series for the BRIC countries\footnote{We have leveraged \say{skforecast} python package to generate Lasso Regression-based forecasts}. The optimal value of the model hyper-parameter ($\alpha$) gets decided through the grid-search cross-validation method for time series.
\item \textit{Xtreme Gradient Boosting} (XGBoost) is a supervised machine learning algorithm that intends to predict the target variable through the ensemble of a set of weak learners \cite{chen2016xgboost}. These tree-based models are trained in a sequential fashion, where the model attempts to minimize the remaining error components of all the prior trees. During prediction, the sum of all the predicted values of the weak learners is reported. In our experimental analysis, we convert the forecasting task into a supervised prediction problem where the lagged inputs and the exogenous variables are the features and the subsequent forecast is the required label for the instance.

\item \textit{Temporal Fusion Transformer} (TFT) is a deep learning time series forecasting model that leverages the multi-head attention layer to capture the complex temporal dynamics of multiple time sequences. One of the key benefits of the TFT framework is its ability to model and forecast both univariate and multivariate time series simultaneously. Furthermore, the architecture is capable of incorporating additional information in the form of time–variant and static exogenous regressors. In our experimentation, we set the number of attention heads as 3.

\item \textit{Wavelet ARIMAx} (WARIMAx) is a modified version of the ARIMAx model that uses the MODWT decomposition technique for the purpose of predicting time series data \cite{aminghafari2007forecasting}. The model breaks down the signal that changes over time into several wavelet and smooth coefficients. It then uses an ARIMAx model to make predictions based on each of these series.The multi-step forecasts are derived by aggregating these candidate forecasts for a desired horizon.

\item \textit{Auto-regressive Neural Network with exogenous variable} (ARNNx) is a generalization of the classical feed-forward neural network model for the auto-regressive time series process \cite{faraway1998time}. This model utilizes previous observations of the target time series and the exogenous variables as input to the network. The ARNNx model comprises a single hidden layer positioned between the input and output layers. The ARNNx $(p, k)$ model uses $p$-lagged input values from the input layers and passes them via $k$ neurons in the hidden layer. The value of $k$ is typically calculated by the formula $k=\left[\frac{p+ 1}{2}\right]$, where $p$ is a given parameter. This approach helps mitigate overfitting and maintain a stable learning architecture \cite{hyndman2018forecasting}. The model gets initialized with random values and trained using the gradient descent back-propagation approach \cite{rumelhart1986learning}. 

\end{itemize}

\subsection{Evaluation Metrics}\label{Sec_evaluation_Metric}
In our study, we have considered five evaluation metrics to assess the performance of the proposed forecaster and the baseline models in generating the 12 months and 24 months ahead forecasts of the CPI inflation series for the BRIC countries. These metrics include the Root Mean Square Error (RMSE), Mean Absolute Scaled Error (MASE), Symmetric Mean Absolute Percentage Error (SMAPE), Theils' U1 measure, Median Relative Absolute Error (MDRAE), and Median Absolute Percentage Error (MDAPE). The mathematical expression of these metrics is provided as follows:
{\scriptsize
\begin{equation*}
 \begin{gathered}
 \text{ RMSE} = \sqrt{\frac{1}{m}\sum_{t=1}^{m} (y_t - \hat{y}_t)^2}; \;
\text{ MASE} = \frac{\sum_{t = D + 1}^{D+m} |\hat{y}_t - y_t|}{\frac{m}{D-S} \sum_{t = S+1}^D |y_t - y_{t-S}|}; \; 
\text{ SMAPE} = \frac{1}{m} \sum_{t=1}^m \frac{|\hat{y}_t - y_t|}{\left(|\hat{y}_t|+ |y_t|\right)/2} \times 100 \% \; \\
\text{ Theils U1} = \frac{\sqrt{\frac{1}{m} \sum_{t = 1}^m (y_t - \hat{y}_t)^2}}{\sqrt{\frac{1}{m} \sum_{t = 1}^m y_t^2} \sqrt{\frac{1}{m} \sum_{t = 1}^m \hat{y}_t^2}}; \; 
\text{ MDRAE} = \underset{t = 1, \ldots, m}{\operatorname{median}} \frac{|y_t - \hat{y}_t|}{y_t - \hat{y}_t^R}; \text{ and } \; 
\text{ MDAPE} = \underset{t = 1, \ldots, m}{\operatorname{median}} \frac{|y_t - \hat{y}_t|}{y_t} \times 100 \%;
\end{gathered}
\end{equation*}
}
where $y_t$ is the ground truth, $\hat{y}_t$ is the forecast of the corresponding model, $\hat{y}_t^R$ is the naive forecast at time $t$, and $m$ is the forecast horizon. By convention, the model with the lowest value of the error metric is considered as the best-performing model \cite{hyndman2018forecasting, panja2023epicasting}.

\subsection{Experimental results and baseline comparison} \label{Sec_setup_result}
This section focuses on the implementation and effectiveness of the proposed FEWNet architecture in forecasting the CPI inflation series of the BRIC countries.  In order to conduct a thorough assessment, we compare the performance of FEWNet to various cutting-edge algorithms. The FEWNet model is implemented using the R statistical software. The MODWT technique is first implemented using the \textit{modwt} function from the `wavelets' package. This function decomposes the training data into wavelet and scaling coefficients using the pyramid algorithm with the `haar' filter. 
The determination of the number of decomposition levels is done conventionally as the floor function of $\log_{e}$(length(train)) \cite{percival2000wavelet}. In the subsequent phase, each wavelet (details) and scaling (smooth) coefficient series is modeled with trend and cycle information from the log-transformed EPU, GPRC, and CPI data obtained through economic filters. This is achieved using an ARNNx model fitted with the \textit{nnetar} function from the `forecast' package in R. To select the number of inputs ($p$) for the individual networks, a time series cross-validation approach is applied with a search range of 1-24. Subsequently, these $p$ inputs are processed through $k = (p+1)/2$ hidden nodes arranged in a single hidden layer. This process generates one-step ahead forecasts for the individual series, which are aggregated to produce the final forecast. The framework iteratively generates multi-step ahead forecasts. The hyperparameters of the FEWNet($p, k$) algorithm used in this study are summarized in Table \ref{FEWNet parameters}. Furthermore, for implementing classical forecasters such as AR, ARIMAx, SARIMAx, ARDL, ARFIMAx, and WARIMAx, the `statsmodels' library in Python is employed. Deep learning models, including DeepAR, NBeats, and TFT, are implemented using Python's `darts' library. The machine learning frameworks like Lasso and XGBoost are implemented using the `skforecast' package in Python. RW, RWD have been implemented using `forecast' package in R statistical software and `msgarch' package in R facilitated fitting the MSGARCH model using the \textit{FitMCMC} function.
\begin{table}[!ht]
\caption{Model parameters: $(p,k)$ of FEWNet for BRIC countries across different forecast – horizons. The table shows optimal parameter combinations for 12 months and 24 months rolling window forecasts of CPI inflation.}
    \centering
    \begin{tabular}{|c|c|c|c|} \hline
    Country & Algorithm & $(p, k)_{FH=12M}$ & $(p, k)_{FH=24M}$ \\ \hline
    Brazil & FEWNet & (6,5) & (18,11) \\ \hline
    Russia  &  FEWNet & (1,2) & (18,11) \\ \hline
    India  &  FEWNet & (18,11) & (12,8) \\ \hline
    China  &  FEWNet & (18,11) & (12,8) \\ \hline
    \end{tabular}
    \label{FEWNet parameters}
\end{table}

After implementing the proposed FEWNet model and other baseline architectures, we generate forecasts for the specified time horizons. Tables \ref{Table_semi_long_term_performance} and \ref{Long_Term_Performance_24M} display the performance of the models for the semi-long-term (12 months) and long-term (24 months) timeframes, respectively, using out-of-sample data. The error metrics reported in the tables indicate that, in the majority of forecasting tasks, the proposed FEWNet framework outperforms the baseline models. In the semi-long-term forecasting of Brazil, the FEWNet model significantly reduces the RMSE metric of the ARNNx model. This improvement in model performance is attributed to the application of wavelet decomposition in the FEWNet model. Other accuracy metrics supports a similar conclusion. 
% For the CPI inflation series of Russia, while the DeepAR model provides the best forecasts among the baseline models, the FEWNet model's forecasts closely align with the ground truth. The 12 months ahead CPI inflation forecasts of India generated by the proposed architecture show substantial improvement over the best-performing NBeats and WARIMAx models. When forecasting the CPI series of China, the NBeats model competes with the FEWNet architecture, with the former exhibiting the best accuracy for all metrics except SMAPE. The Chinese Consumer Price Index (CPI) series demonstrates a pattern of stable trends (trend-stationary). 
For the 12-month forecast horizon, FEWNet consistently outperforms the baseline models for Brazil datasets. In the case of Russia, stochastic AR and FEWNet remained comparative, and their forecasts closely aligned with the ground truth. In the case of India, FEWNet and RWD both outperformed all the benchmark methods by a significant margin. The only stationary time series is the CPI inflation series of China, where the AR model performs the best; however, FEWNet and NBeats remained competitive, and FEWNet obtained the lowest SMAPE amongst its peers. As the Chinese CPI inflation series demonstrates a pattern of steady state (trend-stationary), linear models perform superior to others. While the wavelet-decomposition approach is commendable for tackling structural volatility and non-stationarity, its efficacy may be restricted when applied to trend-stationary or covariance stationary series. Consequently, it may not outperform other statistical or machine-learning methods when used on inherently stationary series (a property that is not prevalent for most macroeconomic variables). This explains the cause of the somewhat competitive or less efficient performance of FEWNet in predicting Chinese inflation numbers. In long-term forecasting, the FEWNet architecture consistently outperforms baseline models for all countries except China and Russia. In the case of China, NBeats, DeepAR, and Lasso regression demonstrate more accurate forecasts of the CPI inflation series. For the Russia CPI inflation series, the AR model outperformed FEWNet and other benchmark methods. Throughout the experimental evaluations, the importance of wavelet transformation and the economically filtered auxiliary information becomes evident in efficiently capturing both time and frequency-level dynamics of the time series. While machine learning and deep learning models generally produce competitive performances for semi-long forecast horizons, their effectiveness diminishes significantly over the long-term horizon. In contrast, the FEWNet model consistently performs well over both horizons, demonstrating its generalizability. Additionally, the wavelet decomposition of the FEWNet approach proves to be a suitable technique for modeling time-frequency information, especially for target variables exhibiting non-stationary behavior (unlike China). Given that most economic time series follow a non-stationary trajectory, the FEWNet approach emerges as an ideal framework for their long-term forecasting even in the presence of structural volatility.

\begin{table*}
\tiny
\centering
\caption{Performance of the proposed FEWNet model in comparison to baseline forecasting techniques for 12 months ahead forecasts with exogenous factors EPU and GPRC (best results are made \textbf{bold}).}
\begin{adjustbox}{width=1\textwidth}
 \begin{tabular}{cccccccccccccccccc} \hline
 Country & Metrics & RW & RWD & AR & MSGARCH & DeepAR & ARNNx & NBeats & ARFIMAx & SARIMAx & ARIMAx & LR & ARDL & XGBoost & TFT & WARIMAx & \textcolor{blue}{Proposed} \\
 &&&&&&&&&&&&&&&&&\textcolor{blue}{FEWNet}\\ \hline
   Brazil & 
   \textit{RMSE} & 4.13 & 4.29	& 3.80 & 3.59 & 3.41 & 2.23 & 2.95 & 4.67 & 2.37 & 2.69 & 2.19 & 3.26 & 2.40 & 5.41 & 3.13 & \textbf{1.30} \\
    & 
   \textit{MASE} & 6.16 &	6.40 &	5.58 &	5.19 & 5.24 & 3.32 & 4.41 & 7.49 & 3.77 & 4.29 & 3.45 & 5.57 & 3.62 & 8.51 & 4.88 &\textbf{1.80} \\
    & 
   \textit{SMAPE(\%)} & 52 &	55 &	45 & 41 &  37 & 25 & 34 & 68 & 29 & 33 & 26 & 52 & 27 & 85 & 39 & \textbf{12} \\
    & 
   \textit{Theil's $U_{1}$} & 0.33 & 0.35 &	0.30	&  0.28 & 0.20 & 0.14 & 0.22 & 0.38 & 0.16 & 0.19 & 0.14 & 0.25 & 0.17 & 0.48 & 0.23 & \textbf{0.09} \\
    & 
   \textit{MDRAE} & 7.18 & 7.44 & 6.59	& 5.36 & 4.42 & 2.45 & 5.22 & 8.38 & 3.84 & 4.44 & 3.70 & 5.24 & 4.41 & 9.26 & 5.45 & \textbf{1.97} \\
    & 
   \textit{MDAPE} & 0.47 & 0.49 & 0.43 & 0.36 & 0.30 & 0.20 & 0.34 & 0.59 & 0.27 & 0.30 & 0.23 & 0.40 & 0.26 & 0.46 & 0.39 & \textbf{0.10} \\ \hline
   %  % &    \textit{MC Ranking} & 8 & 2 & 7 & 11 & 5 & 6 & 3 & 10 & 4 & 12 & 9 & \textbf{1} \\ \hline
   Russia & 
   \textit{RMSE} & 2.14 & 2.49	& 0.99	& 1.14 & 2.00 & 4.16 & 2.49 & 12.68 & 6.65 & 7.80 & 4.45 & 3.65 & 6.21 & 3.87 & 3.72 & \textbf{0.88} \\
    & 
   \textit{MASE} & 5.04 & 5.87	& \textbf{1.95}	& 2.31 & 4.66 & 10.17 & 5.63 & 33.84 & 17.62 & 20.26 & 11.28 & 9.67 & 16.06 & 9.96 & 9.44 & 2.21 \\
    & 
   \textit{SMAPE(\%)} & 33 &	40 & \textbf{11} &	13 & 26 & 46 & 38 & 100 & 69 & 76 & 48 & 44 & 63 & 81 & 44 & 13 \\
    & 
   \textit{Theil's $U_{1}$} & 0.19 & 0.23	& 0.08 & 0.09 & 0.15 & 0.25 & 0.23 & 0.50 & 0.34 & 0.38 & 0.26 & 0.22 & 0.33 & 0.42 & 0.23 & \textbf{0.07} \\
    & 
   \textit{MDRAE} & 4.00 &	4.83 &	\textbf{1.83} &	2.14 & 4.23 & 10.20 & 5.00 & 40.54 & 18.31 & 23.76 & 12.70 & 12.14 & 15.49 & 7.57 & 9.15 & 2.88 \\
    & 
   \textit{MDAPE} & 0.27 &	0.33 &	\textbf{0.08} & 0.12 & 0.33 & 0.51 & 0.32 & 1.97 & 1.05 & 1.22 & 0.71 & 0.57 & 0.99 & 0.59 & 0.56 & 0.11 \\ \hline
   %  % & 
   % % \textit{MC Ranking} & 2 & 6 & 3 & 12 & 10 & 11 & 7 & 5 & 9 & 8 & 4 & \textbf{1} \\ \hline
   India & 
   \textit{RMSE} & 0.99 & 0.91	& 1.23	& 2.18 & 3.55 & 1.69 & 1.09 & 1.86 & 2.27 & 2.31 & 2.46 & 1.88 & 1.69 & 1.67 & 1.19 & \textbf{0.88} \\
    & 
   \textit{MASE} & 1.53 &	\textbf{1.34} &	2.09	& 3.68 & 6.61 & 2.68 & 1.82 & 2.92 & 3.72 & 3.88 & 4.19 & 2.96 & 3.07 & 2.35 & 2.00 & 1.41 \\
    & 
   \textit{SMAPE(\%)} & 16 & \textbf{15} & 21 & 33 & 52 & 33 & 22 & 28 & 51 & 54 & 58 & 38 & 38 & 24 & 23 & \textbf{15} \\
    & 
   \textit{Theil's $U_{1}$} & 0.10 & \textbf{0.09} & 0.12 & 0.19 & 0.27 & 0.21 & 0.12 & 0.17 & 0.29 & 0.30 & 0.33 & 0.23 & 0.21 & 0.16 & 0.13 & \textbf{0.09} \\
    & 
   \textit{MDRAE} & 1.51 &	\textbf{0.81} &	2.66 &	4.90 & 9.18 & 2.18 & 1.70 & 2.43 & 3.91 & 5.16 & 4.26 & 2.37 & 3.93 & 1.86 & 2.46 & 1.22 \\
    & 
   \textit{MDAPE} & 0.15 & \textbf{0.11} & 0.23 &	0.35 & 0.68 & 0.26 & 0.20 & 0.38 & 0.30 & 0.34 & 0.44 & 0.30 & 0.36 & 0.15 & 0.22 & 0.15 \\ \hline
    % & \textit{MC Ranking} & 12 & 5 & 2 & 6 & 9 & 10 & 11 & 7 & 8 & 4 & 3 & \textbf{1} \\ \hline
   China & 
   \textit{RMSE} & 1.58 & 1.71 & \textbf{0.82} & 1.75 & 1.33 & 2.01 & 0.90 & 10.45 & 4.54 & 5.21 & 2.35 & 3.34 & 0.99 & 2.11 & 4.85 & 1.70 \\
    & 
   \textit{MASE} & 2.02 &	2.19 &	\textbf{1.07} &	2.40 & 1.73 & 2.76 & 1.18 & 15.68 & 6.68 & 7.67 & 3.24 & 4.75 & 1.25 & 2.77 & 7.18 & 1.93 \\
    & 
   \textit{SMAPE(\%)} & 173	& 174 & 	103 & 107 & 94 & 111 & 116 & 171 & 146 & 150 & 115 & 132 & 99 & 153 & 149 & \textbf{85} \\
    & 
   \textit{Theil's $U_{1}$} & 0.94 &	0.95	& 0.42	& 0.46 & 0.42 & 0.49 & \textbf{0.34} & 0.83 & 0.68 & 0.71 & 0.53 & 0.62 & 0.35 & 0.80 & 0.70 & 0.43 \\
    & 
   \textit{MDRAE} & 2.28 &	2.44 &	\textbf{1.10} &	2.41 & 1.29 & 2.60 & 1.28 & 15.94 & 7.27 & 8.04 & 3.41 & 5.17 & 1.25 & 3.95 & 7.38 & 1.75 \\
    & 
   \textit{MDAPE} & 1.34 &	1.46 &	\textbf{0.76} &	2.42 & 1.66 & 1.70 & 0.87 & 14.68 & 6.80 & 7.67 & 2.26 & 5.07 & 0.93 & 1.99 & 7.21 & 0.99 \\ \hline
    % &  \textit{MC Ranking} & 3 & 5 & \textbf{1} & 12 & 9 & 11 & 7 & 8 & 2 & 6 & 10 & 4 \\ \hline     
\end{tabular}
\label{Table_semi_long_term_performance}
\end{adjustbox}
\end{table*}

\begin{table*}
\tiny
\centering
\caption{Performance of the proposed FEWNet model in comparison to baseline forecasting techniques for 24 months ahead forecasts with exogenous factors EPU, and GPRC (best results are made \textbf{bold}).}
\begin{adjustbox}{width=1\textwidth}
 \begin{tabular}{cccccccccccccccccc} \hline %{p{0.03\textwidth}p{0.08\textwidth}p{0.038\textwidth}p{0.04\textwidth}p{0.05\textwidth}p{0.05\textwidth}p{0.05\textwidth}ccccccc} \hline %cccccccccccccc
 Country & Metrics & RW & RWD & AR & MSGARCH & DeepAR & ARNNx & NBeats & ARFIMAx & SARIMAx & ARIMAx & LR & ARDL & XGBoost & TFT & WARIMAx & \textcolor{blue}{Proposed} \\
 &&&&&&&&&&&&&&&&&\textcolor{blue}{FEWNet}\\ \hline
 % && \cite{salinas2020deepar} & \cite{faraway1998time} & \cite{oreshkin2019n} && \cite{box1970distribution} & \cite{box1970distribution} && \cite{pesaran1995autoregressive} & \cite{chen2016xgboost} & \cite{lim2021temporal} & \cite{aminghafari2007forecasting} &\textcolor{blue}{FEWNet} \\ \hline
 % \multirow{3}{*}{Brazil} &
   Brazil & 
   \textit{RMSE} & 3.56 & 3.93	& 2.73	& 2.78 & 6.16 & 3.56 & 3.11 & 4.27 & 4.13 & 4.49 & 3.67 & 2.86 & 2.56 & 3.58 & 4.11 &\textbf{2.31} \\
    & 
   \textit{MASE} & 5.34 &	5.89 &	4.34 &	4.64 & 11.45 & 6.68 & 5.10 & 6.04 & 7.67 & 8.27 & 6.93 & 5.23 & \textbf{3.62} & 6.11 & 5.93 & 4.12 \\
    & 
   \textit{SMAPE(\%)} & 50 & 57 & 40 & 42 & 91 & 57 & 47 & 58 & 64 & 69 & 58 & 47 & 40 & 55 & 57 & \textbf{39} \\
    & 
   \textit{Theil's $U_{1}$} & 0.37 & 0.43 & 0.25 & 0.25 & 0.45 & 0.27 & 0.29 & 0.47 & 0.32 & 0.36 & 0.27 & 0.23 & 0.21 & 0.31 &0.45 & \textbf{0.19} \\
    & 
   \textit{MDRAE} & 5.11 & 4.52 & \textbf{2.63} & 2.97 &  12.66 & 6.81 & 5.60 & 3.67 & 9.31 & 8.31 & 7.49 & 5.87 & 5.11 & 6.85 & 3.76 & 4.84 \\
    & 
   \textit{MDAPE} & 0.41 &	0.40 &	0.38 &	0.42 & 0.86 & 0.60 & 0.48 & \textbf{0.36} & 0.56 & 0.65 & 0.70 & 0.44 & 0.41 & 0.50 & 0.39 & \textbf{0.36} \\ \hline
    % &    \textit{MC Ranking} & 12 & 8 & 4 & 7 & 10 & 11 & 9 & 3 & 2 & 5 & 6 & \textbf{1} \\ \hline
   Russia & 
   \textit{RMSE} & 2.06 & 2.74 & \textbf{1.03} & 1.82 & 3.26 & 6.12 & 5.26 & 11.69 & 9.90 & 9.49 & 6.97 & 6.67 & 2.16 & 2.72 &5.22 & 2.24 \\
    & 
   \textit{MASE} & 4.87 & 6.40 & \textbf{2.67} & 4.63 & 8.39 & 18.32 & 14.63 & 35.12 & 28.78 & 27.25 & 20.31 & 19.75 & 5.10 & 7.00 & 15.03 & 5.28 \\
    & 
   \textit{SMAPE(\%)} & 34 &	48  & \textbf{20} &	32 & 51 & 81 & 66 & 112 & 101 & 98 & 84 & 84 & 36 & 44 &72 & 27 \\
    & 
   \textit{Theil's $U_{1}$} & 0.23 &	0.33	& \textbf{0.10} &	0.16 & 0.28 & 0.38 & 0.34 & 0.54 & 0.51 & 0.50 & 0.42 & 0.40 & 0.25 & 0.24 & 0.35 & 0.18 \\
    & 
   \textit{MDRAE} & 4.58 &	5.80 &	\textbf{2.12} &	4.71 & 8.29 & 21.94 & 15.12 & 41.21 & 30.43 & 26.85 & 24.53 & 24.47 & 4.76 & 7.21 & 17.53 & 3.51 \\
    & 
   \textit{MDAPE} & 0.33 &	0.43 &	\textbf{0.20}	& 0.27 & 0.44 & 1.33 & 1.09 & 2.49 & 2.02 & 1.87 & 1.38 & 1.47 & 0.35 & 0.43 & 1.14 & 0.38 \\ \hline
   %  % &    \textit{MC Ranking} & 4 & 7 & 5 & 12 & 11 & 10 & 9 & 8 & 2 & 3 & 6 & \textbf{1} \\ \hline
   India & 
   \textit{RMSE} & 3.72 & 3.73 & 2.92 & 1.92 & 1.86 & 2.22 & 4.18 & 2.72 & 2.16 & 2.02 & 2.36 & 2.53 & 1.95 & 6.17 & 2.35 & \textbf{1.04} \\
    & 
   \textit{MASE} & 6.20 & 6.21 & 4.87 & 2.77 & 2.68 & 2.90 & 5.83 & 4.20 & 3.02 & 2.83 & 3.61 & 2.98 & 2.80 & 9.48 & 3.39 & \textbf{1.49} \\
    & 
   \textit{SMAPE(\%)} & 51 & 51 & 43 & 26 & 26 & 35 & 46 & 38 & 34 & 31 & 47 & 37 & 33 & 65 & 37 & \textbf{17} \\
    & 
   \textit{Theil's $U_{1}$} & 0.26 & 0.26 & 0.21	& 0.15 & 0.15 & 0.23 & 0.29 & 0.21 & 0.21 & 0.19 & 0.26 & 0.26 & 0.20 & 0.37 & 0.21 & \textbf{0.10} \\
    & 
   \textit{MDRAE} & 10.76 &	10.80 & 8.25 & 3.71 & 3.25 & 4.31 & 6.22 & 6.46 & 4.02 & 4.07 & 4.21 & 2.12 & 4.28 & 13.47 & 3.70 & \textbf{1.42} \\
    & 
   \textit{MDAPE} & 0.70 & 0.70	& 0.54	& 0.24 & 0.31 & 0.24 & 0.49 & 0.42 & 0.30 & 0.28 & 0.41 & 0.21 & 0.26 & 1.07 & 0.40 & \textbf{0.14} \\ \hline
   %  % &     \textit{MC Ranking} & 2 & 5 & 11 & 10 & 6 & 3 & 9 & 7 & 4 & 12 & 8 & \textbf{1} \\ \hline
   China & 
   \textit{RMSE} & 3.15 &	3.23 &	2.23 & 2.08 & 1.71 & 1.85 & \textbf{1.43} & 18.37 & 3.48 & 4.04 & 1.59 & 2.00 & 2.87 & 2.05 & 7.40 & 2.01 \\
    & 
   \textit{MASE} & 3.95 & 4.06 & 2.74 & 2.63 & 1.95 & 2.15 & \textbf{1.75} & 25.72 & 4.23 & 5.05 & 1.89 & 2.33 & 3.12 & 2.42 & 10.17 & 2.37 \\
    & 
   \textit{SMAPE(\%)} & 101 & 102 & 89 & 89 & 81 & 77 & 77 & 168 & 102 & 109 & \textbf{75} & 79 & 115 & 100 & 139 & 82 \\
    & 
   \textit{Theil's $U_{1}$} & 0.45 &	0.46 &	0.36 &	0.35 & 0.40 & 0.34 & \textbf{0.27} & 0.81 & 0.48 & 0.51 & 0.30 & 0.34 & 0.65 & 0.46 & 0.64 &0.36 \\
    & 
   \textit{MDRAE} & 3.87 & 4.07 & 2.61 & 2.42 & \textbf{1.53} & 2.32 & 1.60 & 26.02 & 4.43 & 4.99 & 2.09 & 2.51 & 2.68 & 2.32 & 10.88 & 2.50 \\
    & 
   \textit{MDAPE} & 1.68 & 1.73 & 1.14 & 0.95 & 0.57 & 0.85 & 0.63 & 11.56 & 2.05 & 2.41 & \textbf{0.55} & 0.99 & 1.14 &0.84 & 4.84 & 0.84 \\ \hline
    % &    \textit{MC Ranking} & 3 & 6 & \textbf{1} & 12 & 9 & 10 & 7 & 8 & 2 & 5 & 11 & 4 \\ \hline    
\end{tabular}
\label{Long_Term_Performance_24M} 
\end{adjustbox}
\end{table*}

\subsection{Statistical significance of the results}\label{Sec_stat_signif}
\begin{figure}[h!]
 \centering
  \includegraphics[width=0.75\textwidth]{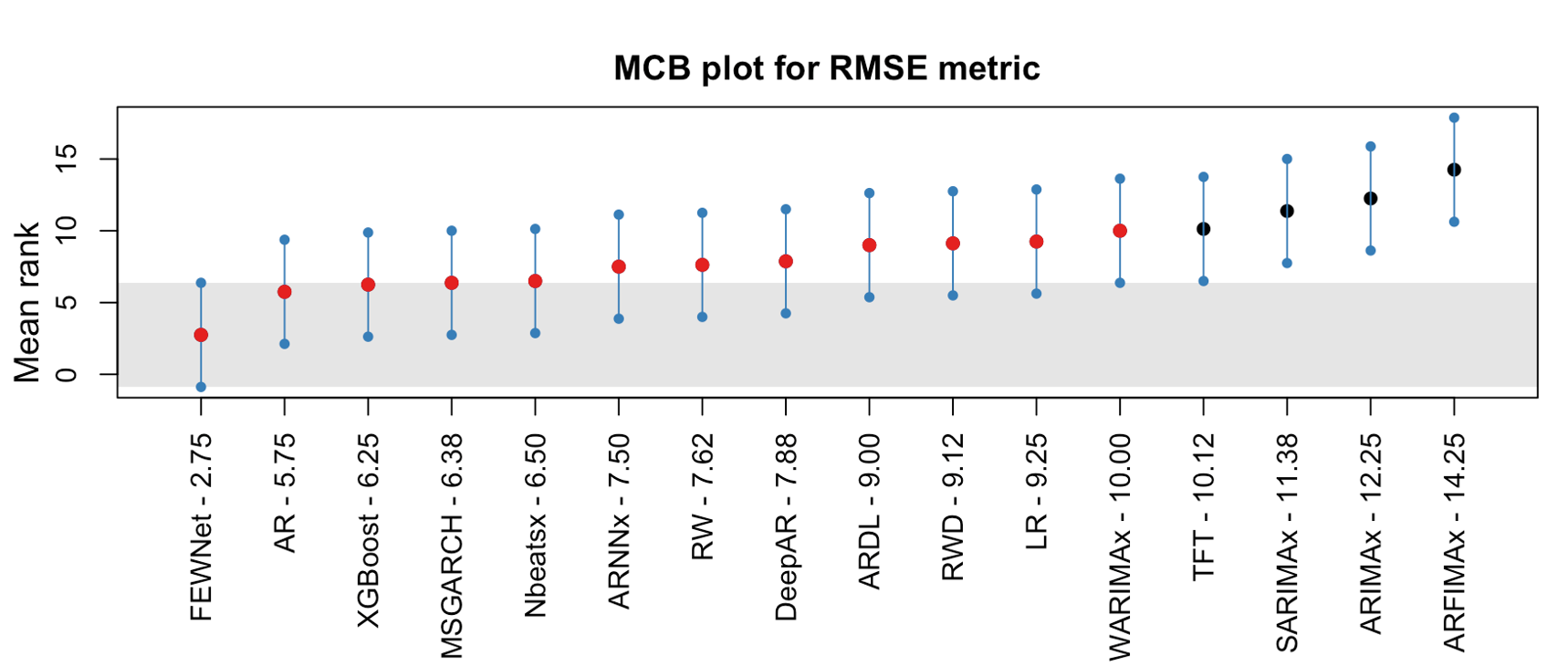}
   \caption{Visualization of the multiple comparisons with best (MCB) analysis for BRIC countries. In the figure, for example, FEWNet - 2.75 means that the average rank of the proposed algorithm FEWNet based on the RMSE error metric is 2.75; the same explanation applies to other algorithms.}
\label{fig:MCB_test}
\end{figure}

In this section, we assess the effectiveness of different forecasting models in terms of the statistical significance of measurement error differences using the model-agnostic multiple comparisons with the best (MCB) procedure \cite{koning2005m3}. This non-parametric test computes the average rank for each of the $\mathcal{M}$ forecasters based on their ex-ante accuracy across $\mathcal{D}$ datasets and identifies the model with the minimum average rank as the `best' performing technique. The MCB test determines critical distances (CD) for each algorithm as $\Theta_{\alpha} \sqrt{\frac{\mathcal{M}\left(\mathcal{M} + 1 \right)}{6 \mathcal{D}}}$, where $\Theta_{\alpha}$ is the critical value of the Tukey distribution at level $\alpha$, and it considers the CD of the best-performing model as the reference value for the test.
Following the MCB test, we calculate the algorithm's average rank and present the MCB test results for the RMSE metric in Figure \ref{fig:MCB_test}. The plot reveals that the FEWNet framework has achieved the lowest mean rank of 2.75 among all competing algorithms for both forecast horizons. Therefore, it is deemed to be the best-performing model, followed by AR (5.75), XGBoost (6.25), MSGARCH (6.38), NBeats (6.50), and so forth.
% Moreover, the upper boundary of the critical distance for the FEWNet model (indicated by the greyed-out region in the plots) serves as the reference value for the test. 
Furthermore, the upper limit of the critical distance for the FEWNet model (shown by the shaded area in the graphs) acts as the reference value for the test. 
% When compared to the baseline, SARIMAx, ARIMAx, and ARFIMAx exhibit critical intervals well above the reference value without overlap with FEWNet, signifying that their forecast performance is significantly inferior to the FEWNet method.
When comparing SARIMAx, ARIMAx, and ARFIMAx to the baseline, it is observed that their critical intervals are much higher than the reference value and do not coincide with FEWNet. This indicates that their forecast performance is notably worse than the FEWNet approach. AR model obtained second place and fell short of FEWNet's performance. MSGARCH, although designed for handling volatility, does not surpass FEWNet. Our analysis underscores the statistical significance of the performance difference and the superiority of the proposed FEWNet model across datasets and forecast horizons. Moreover, the test establishes that FEWNet consistently and significantly outperforms other baseline models, as well as various machine learning or deep learning models.

\begin{figure}
    \centering
    \includegraphics[width=0.85\textwidth]{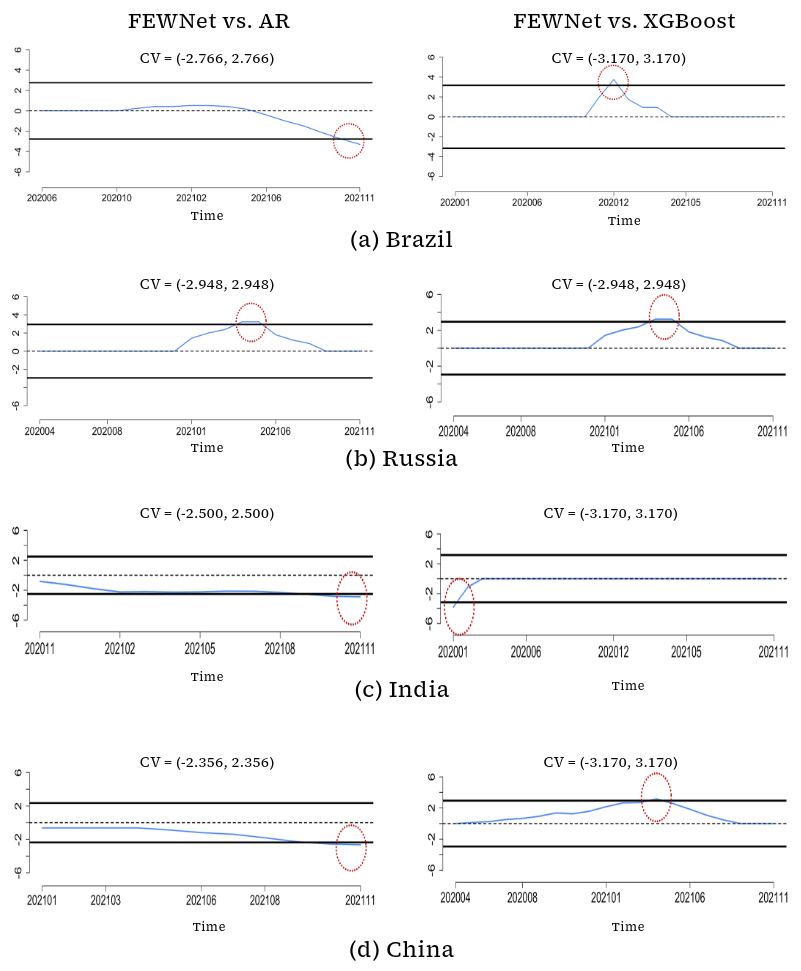}
    \caption{Results of the GR test (Giacomini \& Rossi Test) of forecasting ability of the FEWNet with baselines (AR (left) and XGBoost (right)) with ($\alpha$=0.90 and $\mu$=1) for (a) Brazil, (b) Russia, (c) India, and (d) China. Points outside the CV have been marked under the red dotted circles representing that FEWNet and other two top performing benchmarks differ significantly.}
    \label{fig:GR_test_FEWNet}
\end{figure}

In addition to the MCB test, the Giacomini and Rossi (GR) test \cite{giacomini2010forecast} is employed to assess the robustness and forecasting ability of the proposed FEWNet model under ``unstable environments’’ in comparison to other baseline models. The GR test is designed to evaluate whether one forecasting model consistently outperforms another over a specified period, particularly in data with structural breaks or regime shifts, which traditional tests like the Diebold-Mariano test might not effectively handle \cite{diebold2002comparing, rossi2016forecast}. GR test specifically examines the stability of relative local forecasting performance over time. We used the `murphydiagram’ package in R to perform the GR test. Based on the MCB test results, the top two baseline models identified are AR and XGBoost. We employ the GR test to further evaluate the stability of the relative predictive accuracy of these models over time. This test facilitates pairwise comparisons between FEWNet and each of the two baseline models, enabling a rigorous assessment of their relative performance. GR test is performed at a 10\% level of significance ($\alpha$ = 0.90), with critical value (CV) thresholds indicated by thick black line in the plots. The fluctuation test diagrams display varying widths of CV bounds and different endpoints for the rolling window across time. This variability can be attributed to the choice of the parameter ($\mu$), which captures the size of the rolling window relative to the evaluation sample during GR test implementation. The null hypothesis of the fluctuation test (GR test) posits that both forecasting methods perform equally well (same expected score) at all time points, while the alternative hypothesis seeks to identify if their performance differs at any time point. The objective is to find the value of $\mu$ that helps pinpoint where the performance divergence occurs. Consequently, different forecasting methods result in different CV thresholds and rolling window sizes, reflecting the unique points of performance variation over time. These charts are based on long-term (24 months) forecasting horizons (refer to Figure \ref{fig:GR_test_FEWNet}). From the GR test, we can conclude that the proposed FEWNet model consistently outperforms the top-performing baseline models (AR and XGBoost) across the BRIC countries during test periods. These findings collectively establish FEWNet as a more accurate and stable forecasting model under dynamic economic conditions, highlighting its potential as a valuable tool for economic forecasting and policy analysis.

\subsection{Empirical results on ERM}\label{Sec_empirical_validation_of_results}
We investigate the empirical risk of the proposed FEWNet model on BRIC datasets (training series). Empirically, ERM helps to improve the model's accuracy by minimizing the empirical risk and reducing the level of error or deviation between the predicted and actual values of the dataset. The key challenge in the inflation series is to learn fast dynamics and to cancel the noise simultaneously. We tackle this problem by iterative application of wavelet decomposition using low and band-pass filters, resulting in a trend series and a hierarchy of detail series that contains the system's dynamics at different scales. Several ARNNx are trained to model these decomposed series and provide forecasts of the original time series by combining individual forecasts. We theoretically have proved in Section \ref{ERM_FEWNet} that using the FEWNet method (incorporated wavelet decomposition) reduces the empirical risk under some conditions. We now carry out practical experiments on BRIC countries' datasets to show that the use of the wavelet coefficients is robust to model estimation and reduces the generalization risk for the majority of the training samples as depicted in Table \ref{ERM_Practical_Implications}. It is clear from Table \ref{ERM_Practical_Implications} (in-sample predictions) that FEWNet has improved performance for highly non-stationary time series as compared to standard neural networks and other benchmarks also for out-of-sample forecasts as depicted in Tables \ref{Table_semi_long_term_performance} and \ref{Long_Term_Performance_24M}. This has a series of practical implications and is important in policymaking. In an ideal situation, FEWNet will always perform better than the autoregressive neural network model. Also, as FEWNet can separate the fast dynamics from the slow ones and generate a combined forecast, it escapes the problem of underfitting and overfitting. Thus, FEWNet is more suitable for countries with complex, non-stationary CPI inflation time series and will be very useful for policymakers at central banks in BRIC countries to easily visualize the future path of inflation. A more detailed discussion on policy implications is given in Section \ref{Sec_Policy_Implications}.

\begin{table*}
\centering
\caption{Empirical risk comparison of two different training periods (best results are made \textbf{bold}).}
\scriptsize
    \centering
    \begin{tabular}{cccccc}
\hline Training Period & Models & Brazil & Russia & India &  China \\
\hline 2003-01 to 2020-11 & ARNNx & 2.61 & 16.62 & \textbf{0.51} & 1.16 \\
                          & FEWNet & \textbf{1.14} & \textbf{0.77} & 1.38 & \textbf{1.10} \\ \hline
       2003-01 to 2019-11 & ARNNx & 5.79 & 17.66 & \textbf{0.54} & 1.23 \\
                          & FEWNet & \textbf{4.60} & \textbf{13.13} & 0.93 & \textbf{1.01} \\ \hline
    \end{tabular}
        \label{ERM_Practical_Implications}
\end{table*}

\subsection{Robustness and sensitivity analysis}\label{sec_robustness_sensitivity}

This section provides a more comprehensive robustness and sensitivity analysis of the FEWNet method. For instance, we first study the impact of HP and CF filters in the FEWNet model. This is of particular interest since \cite{hamilton2018you} criticized the HP filter for generating spurious dynamic relations and producing end-sample values that differ significantly from mid-sample values. \cite{drehmann2018you} rejected this criticism by highlighting scenarios in macroeconomic contexts where the HP filter is useful and robust. HP filter is more accurate when applied to medium and long-term cycles instead of short-term fluctuations. We employ an empirical approach to verify the importance of HP and CF economic filters in the FEWNet model. HP filter is used to extract trend components, whereas CF filter is employed to isolate cyclical components from the CPI inflation, EPU, and GPRC series. We examined the relevance of HP and CF filters by comparing the experimental results of the FEWNet model with and without economic filters. Tables \ref{tab:FEWNet_Versions_12M_Comp} and \ref{tab:FEWNet_Versions_24M_Comp} summarize the performance of FEWNet and FEWNet without HP and CF filters (referred to as EWNet) for semi-long-run and long-run. The results indicate that these economic filters play a critical role in CPI inflation series analysis. The tables underscore the substantial improvement in predictive accuracy and robustness of the FEWNet model when economic filters are applied. This finding highlights the importance of integrating trend and cycle components using HP and CF filters, which enhances the model’s ability to capture the underlying dynamics of economic series. Consequently, the FEWNet model with economic filters proves to be a more effective tool for long-term economic forecasting in the BRIC countries, addressing and overcoming the traditional criticisms of using the HP filter.

In addition, we analyze the sensitivity of the experimental results relative to the choice of different wavelet filters and the level of wavelet decomposition. The selection of wavelet filters is crucial in the overarching design of the proposed FEWNet architecture. The model employs the ``haar'' wavelet transform filter to decompose the underlying CPI inflation series into detailed and smooth components. In order to rationalize the choice of the wavelet transform filter, we conduct an iterative out-of-sample forecast evaluation, comparing the performance of these competing models (FEWNet Model with different wavelet filters) over the semi-long-term (12 months) and long-term (24 months) forecast horizon. A grid-search approach is used to determine the optimal set of hyperparameters for the underlying ARNNx model within the FEWNet framework. The selection of the wavelet decomposition level ($\mathcal{K}$) is determined by the formula $\mathcal{K} = \lfloor \log_e(\mathcal{N}) \rfloor$, where $\mathcal{N}$ represents the size of the training observations \cite{percival2000wavelet}. This calculation yields $\mathcal{K}=5$. The same decomposition level is applied uniformly to ensure a fair comparison among different wavelet filters. It is standard practice in financial forecasting exercises to utilize 5 to 6 levels of detail. For monthly data, the first detail ($\delta_1$) captures oscillations between 2-4 months, $\delta_2$ between 4-8 months, $\delta_3$ between 8-16 months, $\delta_4$  between 16-32 months, and $\delta_5$ between 32-64 months. This approach effectively captures business cycle frequency dynamics, as additional details would require longer time period information, which is often unavailable for most macroeconomic series \cite{souropanis2023forecasting}. For a comprehensive comparison, we have considered four other wavelet transform filters, such as (a) Daubechies with length 8 (d8): known for its orthogonality and compact support, the Daubechies wavelet is effective in capturing intricate structures in the data; (b) Least asymmetric with length 8 (la8): this wavelet balances asymmetry and orthogonality, providing a more symmetric filter compared to Daubechies; (c) Coiflet with length 6 (c6): the Coiflet wavelet emphasizes both orthogonality and symmetry, making it suitable for analysis requiring smooth approximation components; (d) Best localized with length 14 (bl14): this wavelet is designed for optimal localization in both time and frequency domains, enhancing the detection of localized features in the data. The decomposition levels are uniformly set to five for all wavelet transform filters, utilizing the MODWT-based MRA approach.

\begin{figure}[h!]
 \centering
  \includegraphics[width=0.75\textwidth]{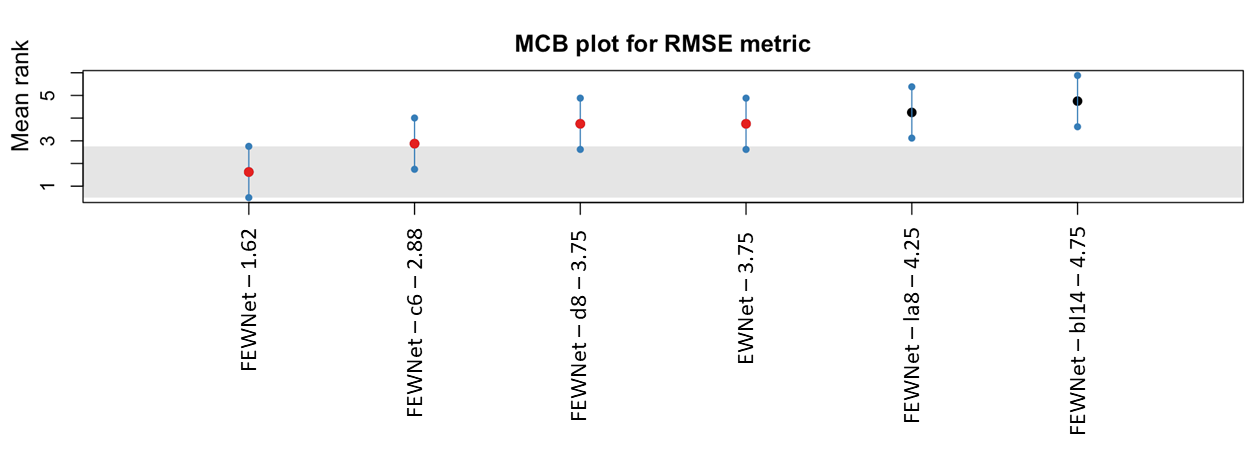}
   \caption{Visualization of the MCB analysis for BRIC countries for different FEWNet variants of the wavelet filters along with the proposed FEWNet algorithm. In the figure, for example, FEWNet - 1.62 means that the average rank of the FEWNet based on the RMSE error metric is 1.62; the same explanation applies to its variants based on different wavelet filters.}
\label{fig:MCB_test_Variant}
\end{figure}

Figure \ref{fig:MCB_test_Variant} showcases that the FEWNet model with the `haar' wavelet filter consistently achieves the lowest mean rank across the BRIC countries, indicating its best-in-class performance among the evaluated wavelet transform filters. The rationale for the `haar' wavelet's superior performance lies in its simplicity and capability to handle abrupt changes in data. Unlike more complex wavelets, the `haar' wavelet's structure allows it to efficiently manage and adapt to the volatility inherent in economic series, particularly in the presence of regime shifts or exogenous shocks. This efficiency is reflected in its lower forecast errors and greater stability in predictions.

\begin{table}[]
\centering
\caption{Performance comparison between the proposed FEWNet model and their variants for 12 months ahead forecasts with exogenous factors EPU and GPRC (best results are made bold).}
\tiny
    \centering
    \begin{tabular}{cccccccc}
\hline Country & Metrics & FEWNet & FEWNet - d8 & FEWNet - la8 &  FEWNet - c6 & FEWNet - bl14 & EWNet \\
\hline 
Brazil & RMSE & \textbf{1.30} & 2.76 & 3.10 & 1.36 & 1.93 & 3.63 \\
       & MASE & 1.80 & 4.16 & 4.79 & \textbf{1.77} & 2.70 & 5.94 \\
       & SMAPE(\%) & \textbf{12} & 32 & 39 & \textbf{12} & 20 & 54 \\
       & Theil's $\mathrm{U}_1$ & \textbf{0.09} & 0.19 & 0.22 & \textbf{0.09} & 0.13 & 0.28 \\
       & MDRAE & 1.97 & 3.81 & 7.22 & \textbf{0.92} & 3.50 & 5.29 \\
       & MDAPE & \textbf{0.10} & 0.27 & 0.36 & 0.11 & 0.18 & 0.44 \\ \hline
       
Russia & RMSE & 0.88 & 2.47 & 0.85 & \textbf{0.57} & 5.83 & 2.49 \\
       & MASE & 2.21 & 5.90 & 2.16 & \textbf{1.29} & 15.48 & 4.88 \\
       & SMAPE(\%) & 13 & 40 & 13 & \textbf{8} & 63 & 32 \\
       & Theil's $U_1$ & 0.07 & 0.23 & 0.06 & \textbf{0.04} & 0.31 & 0.22 \\
       & MDRAE & 2.88 & 4.99 & 2.13 & \textbf{1.24} & 16.64 & 3.42 \\
       & MDAPE & 0.11 & 0.29 & 0.14 & \textbf{0.07} & 1.00 & 0.21 \\ \hline
       
India & RMSE & \textbf{0.88} & 1.41 & 2.73 & 2.15 & 3.59 & 1.77 \\
      & MASE & \textbf{1.41} & 2.45 & 5.14 & 3.48 & 6.62 & 3.34 \\
      & SMAPE(\%) & \textbf{15} & 28 & 44 & 31 & 53 & 31 \\
      & Theil's $U_1$ & \textbf{0.09} & 0.15 & 0.22 & 0.19 & 0.28 & 0.16 \\
      & MDRAE & \textbf{1.22} & 3.28 & 6.18 & 2.49 & 7.79 & 4.61 \\
      & MDAPE & \textbf{0.15} & 0.24 & 0.57 & 0.34 & 0.74 & 0.36 \\ \hline
      
China & RMSE & 1.70 & 4.06 & 1.85 & 2.39 & 4.63 & \textbf{0.86} \\
      & MASE & 1.93 & 5.87 & 2.06 & 3.47 & 6.83 & \textbf{1.12} \\
      & SMAPE(\%) & \textbf{85} & 138 & 91 & 123 & 148 & 105 \\
      & Theil's $U_1$ & 0.43 & 0.67 & 0.54 & 0.53 & 0.69 & \textbf{0.36} \\
      & MDRAE & 1.75 & 7.06 & 1.16 & 3.57 & 6.20 & \textbf{1.00} \\
      & MDAPE & \textbf{0.99} & 5.23 & 1.62 & 3.61 & 7.08 & 1.04 \\ \hline
\end{tabular}
    \label{tab:FEWNet_Versions_12M_Comp}
\end{table}

\begin{table}[]
\centering
\caption{Performance comparison between the proposed FEWNet model and their variants for 24 months ahead forecasts with exogenous factors EPU and GPRC (best results are made bold).}
    \tiny
    \centering
    \begin{tabular}{cccccccc}
\hline Country & Metrics & FEWNet & FEWNet - d8 & FEWNet - la8 &  FEWNet - c6 & FEWNet - bl14 & EWNet \\
\hline 
Brazil  & RMSE & 2.31 & \textbf{1.92} & 4.06 & 3.42 & 2.82 & 2.65 \\
     & MASE & 4.12 & \textbf{3.30} & 7.59 & 6.04 & 4.66 & 4.52 \\
     & SMAPE(\%) & 39 & \textbf{34} & 64 & 52 & 43 & 41 \\
     & Theil's $\mathrm{U}_1$ & 0.19 & \textbf{0.15} & 0.31 & 0.24 & 0.25 & 0.23 \\
     & MDRAE & 4.84 & \textbf{3.14} & 9.31 & 6.12 & 5.51 & 5.12 \\
     & MDAPE & 0.36 & \textbf{0.22} & 0.71 & 0.65 & 0.40 & 0.41 \\ \hline 

Russia  & RMSE & \textbf{2.24} & 4.84 & 2.94 & 3.56 & 3.81 & 2.90 \\
     & MASE & \textbf{5.28} & 12.96 & 7.09 & 9.53 & 9.92 & 6.71 \\
     & SMAPE(\%) & \textbf{27} & 116 & 40 & 77 & 51 & 41 \\
     & Theil's $\mathrm{U}_1$ & \textbf{0.18} & 0.55 & 0.24 & 0.39 & 0.28 & 0.23 \\
     & MDRAE & \textbf{3.51} & 12.46 & 6.29 & 11.08 & 6.88 & 7.61 \\
     & MDAPE & \textbf{0.38} & 0.87 & 0.36 & 0.72 & 0.52 & 0.45 \\ \hline

India & RMSE & \textbf{1.04} & 2.32 & 3.57 & 1.96 & 3.44 & 2.35 \\
      & MASE & \textbf{1.49} & 3.20 & 5.78 & 2.87 & 5.03 & 2.88 \\
     & SMAPE(\%) & \textbf{17} & 44 & 48 & 26 & 43 & 26 \\
     & Theil's $\mathrm{U}_1$ & \textbf{0.10} & 0.25 & 0.25 & 0.16 & 0.25 & 0.19 \\
     & MDRAE & \textbf{1.42} & 3.55 & 9.58 & 3.85 & 6.44 & 3.24 \\
     & MDAPE & \textbf{0.14} & 0.28 & 0.59 & 0.30 & 0.48 & 0.20 \\ \hline
     
China & RMSE & 2.01 & 2.43 & 2.32 & \textbf{1.90} & 2.22 & 3.76 \\
     & MASE & 2.37 & 3.00 & 2.75 & \textbf{2.14} & 2.68 & 4.69 \\
     & SMAPE(\%) & \textbf{82} & 97 & 85 & 98 & 87 & 106 \\
     & Theil's $\mathrm{U}_1$ & \textbf{0.36} & 0.45 & \textbf{0.36} & 0.37 & 0.37 & 0.49 \\
     & MDRAE & 2.50 & 2.46 & 2.83 &\textbf{ 1.86} & 2.37 & 5.06 \\
     & MDAPE & 0.84 & 0.99 & 0.91 & \textbf{0.71 }& 1.07 & 2.18 \\
\hline
\end{tabular}
    \label{tab:FEWNet_Versions_24M_Comp}
\end{table}

\subsection{Conformalized Prediction Intervals}\label{Sec_conf_pred}
\begin{figure}[h!]
 \centering
  \includegraphics[width=\textwidth]{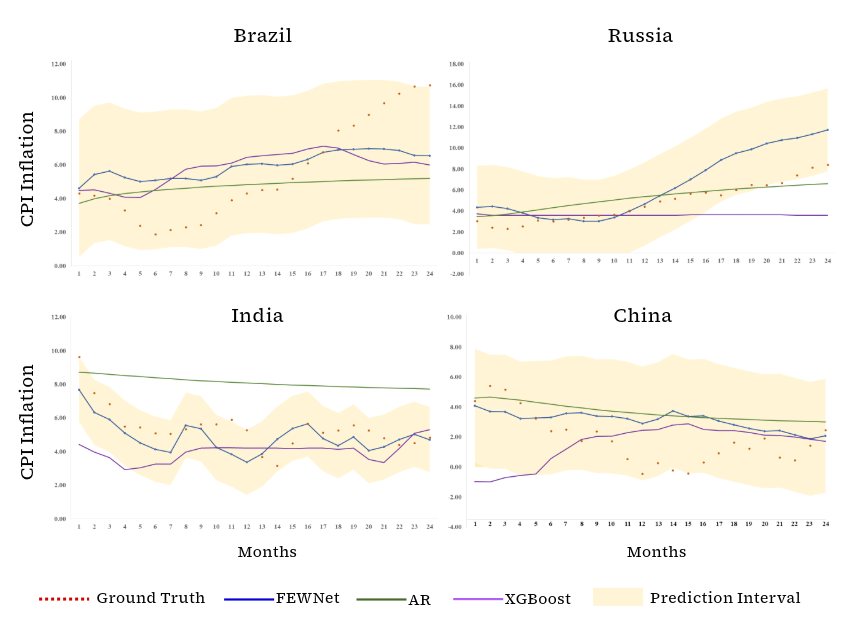}
   \caption{The plot shows the ground truth CPI inflation data (red dots), 24 months point forecasts generated by FEWNet (blue line), AR (green line), XGBoost (purple line), and the conformal prediction interval of FEWNet (yellow shaded) for BRIC Countries.}
   \label{fig:CP_long}
\end{figure}

Next, we focus on quantifying the uncertainty associated with the forecast generated by the FEWNet approach using a conformal prediction interval. The non-parametric conformal prediction approach, introduced in \cite{vovk2005algorithmic}, is a set of methods that usually take an uncertainty score and turn it into a probabilistic band, including the true outcome. In other words, this framework converts point estimations into a prediction region. One of the significant advantages of conformal prediction intervals is that this model-agnostic procedure guarantees coverage. In the time series setup, this method generates the prediction interval by leveraging the sequential ordering of the data. Given the training set $\left\{Y_t, \tilde{X}_t \right\}_{t=1}^N$ where $Y_t$ is the target series, and $\tilde{X}_t$ is the set of features, including lagged historical values of $Y_t$ and the covariates, we fit the FEWNet framework and an uncertainty model $\widehat{\Xi}$ on $\tilde{X}_t$ to generate a scaler notion of uncertainty. The conformal score $\mathcal{S}_{t}$ can be calculated as:
\begin{equation}
    \mathcal{S}_{t} = \frac{\left|Y_{t} - \operatorname{FEWNet}\left(\tilde{X}_{t} \right)\right|}{\widehat{\Xi}(\tilde{X}_{t})}
\end{equation}
Owing to the sequential nature of the data points, the conformal quantile for time series data can be calculated as a weighted conformal technique with a fixed $\kappa$-sized window $\omega_{t^{'}} = \mathbb{1}\{ t^{'} \geq t - \ \kappa), \forall\ t^{'} < t$. This now yields the quantile values as,
\begin{equation}
\widehat{\mathbb{Q}}_{t} = inf\left\{ q:\ \frac{1}{\min\left( \kappa,t^{'} - 1 \right) + 1}\sum_{t^{'} = 1}^{t - 1}{\mathcal{S}_{t^{'}}\mathbb{1}\left(t^{'} \geq t - \ \kappa \right) \geq 1 - \alpha}\right\}
\end{equation}
Based on these weights-adjusted quantiles, the prediction interval at each time step $t$ can be computed as,
\begin{equation}
\mathbb{C}\left(\tilde{X}_{t} \right) = \left[\operatorname{FEWNet}\left(\tilde{X}_{t} \right) - {\widehat{\mathbb{Q}}}_{t} \widehat{\Xi}\left(\tilde{X}_{t} \right), \operatorname{FEWNet}\left( \tilde{X}_{t} \right) + {\widehat{\mathbb{Q}}}_{t}\ \widehat{\Xi}\left(\tilde{X}_{t} \right)\right].
\end{equation}
In this section, we present the conformal prediction intervals for the 24 months ahead of CPI inflation forecasts generated by the FEWNet model using the validation data points (2019-12 to 2021-11). The conformal prediction interval of the FEWNet framework, calculated based on the forecast estimates and residuals for the validation set, is demonstrated in Figure \ref{fig:CP_long} along with the point estimate of the forecast produced by FEWNet and the best-performing baseline models for the BRIC countries. Although the AR model ranked $2^{nd}$ based on the performance metrics however, it significantly fails to capture the movements in the test CPI inflation series and may not be an effective forecasting strategy. As observed in the figure, the prediction intervals vary in width across different datasets. For instance, in the case of the 24-month forecast for Brazil, the width of the interval averages around 8.18; for Russia, it hovers around 7.93; for India, it reaches the minimum of 3.88; and for China, it averages around 7.60. A similar analysis of the conformal prediction intervals for the semi-long-term forecast horizon is provided in Appendix \ref{App_CP_SL}. The overall analysis thus demonstrates a form of uncertainty quantification for the forecast of CPI inflation series for BRIC countries and shows the structural shift in the measurement of prediction uncertainty across different countries.

\section{Policy Implications}\label{Sec_Policy_Implications}

Most of the BRIC countries, except China, have adopted inflation targeting as a framework for monetary policy, in which inflation serves as an intermediate target. Accurately forecasting of inflation in such countries is essential for maintaining the credibility of these targets. By reliably forecasting inflation, central bankers in these countries can communicate their policy intentions, manage public expectations, and build trust in their commitment to price stability. However, as can be seen from Tables \ref{Table_Global characteristics} and \ref{PLOTS:ACF}, inflation trends in BRIC countries are characterized by complex, volatile, and non-linear patterns influenced by numerous internal and external factors. Two major factors that lead to such patterns are EPU and GPRC, which affect the demand or supply side of the economy \cite{adeosun2023uncertainty}, leading to large fluctuations in inflation \cite{balcilar2017long}. For example, BRIC countries depend significantly on commodities such as oil, gas, and agricultural products, among many others. Consequently, fluctuations in global commodity prices due to geopolitical risk can directly impact their inflation rates. Additionally, EPU can cause increased volatility in financial markets, affecting exchange rates and commodity prices. For BRIC countries, which are often significant commodity exporters, this can lead to fluctuating import and export prices, impacting inflation. Moreover, geopolitical tensions (e.g., the Russia-Ukraine War) can lead to economic sanctions, trade disruptions, and shifts in global economic alliances, further affecting inflation through these channels. From the central banks' point of view, high EPU and GPRC make it harder for central banks to set appropriate interest rates. Reactive monetary policies may become less effective, leading to more pronounced inflation volatility and making it harder to maintain price stability \cite{balcilar2017long}. Given the dynamic relation between inflation, EPU, and GPRC in BRIC countries, it is essential to incorporate these uncertainty measures as exogenous variables in inflation forecasting models. Ignoring them can make the forecasting process even more challenging and result in inaccuracies for central banks \cite{binder2024central, nasir2020forecasting}. Traditional forecasting models often struggle to adapt to sudden changes and capture the non-linear dynamics characteristic of these environments. In contrast, machine learning models, particularly ensemble models, offer significant advantages for inflation forecasting in uncertain times due to their ability to process vast amounts of data and identify complex patterns \cite{john2020inflation, medeiros2021forecasting}.

In this paper, by employing the wavelet coherence analysis, we show a significant causal relation between inflation, EPU, and GPRC in BRIC countries. Excluding them in the inflation forecasting models will lead to model misspecification, resulting in incorrect forecasts\footnote{For instance, during both the global financial crisis and the COVID-19 pandemic, central banks in advanced economies experienced significant forecasting errors, notably underestimating inflation \cite{alessi2014central}. Researchers at the European Central Bank identified a major cause for these errors: the neglect of various financial and uncertainty indicators and the failure to account for non-linear dynamics in forecasting models \cite{kenny2011some}.}. Because the transmission channels of inflation and monetary policy during periods of high EPU (such as the global financial crisis 2007-09) and GPRC change substantially \cite{benchimol2016money}. This argument is further supported by the findings of \cite{benchimol2020forecast} where they conclude that while predicting inflation for a country like Israel, the policymakers should consider the current level of terrorism (one of the aspects of GPRC) in their inflation forecasting model. 

Furthermore, we show that our proposed model, FEWNet, performed superior in forecasting CPI inflation for Brazil, Russia, and India for semi-long-run and long-run horizons. FEWNet integrates economic filters like HP and CF with the wavelet decomposed autoregressive neural networks to significantly improve the forecast accuracy compared to traditional models. Specifically, we can conclude that FEWNet is more suitable for countries where inflation is characterized by non-stationarity and non-linearity due to heightened volatility and other complex dynamics. Given these findings, we recommend that when policymakers at central banks in BRIC countries evaluate the future path of inflation, they should include uncertainty and risk variables like EPU and GPRC and adopt advanced machine learning models like the FEWNet model for more accurate inflation forecasting. This is also evident from the theory (Section \ref{ERM_FEWNet}) and empirical study (Section \ref{Sec_empirical_validation_of_results}) of the empirical risk for the proposed FEWNet model. With more accurate inflation forecasts, central banks in these countries can better align their monetary policies with actual economic conditions. This reduces the risk of policy errors, such as setting interest rates that are too high or too low relative to the economic context. Advanced models like FEWNet enable central banks to more effectively anticipate and respond to inflationary trends. This proactive approach facilitates timely and targeted interventions, especially in countries with complex and dynamic economic conditions, helping to stabilize inflation expectations and maintain economic stability.

\section{Conclusion and Discussion}\label{Section_Conclusion}
%Historically, emerging economies have exhibited higher and more enduring inflation rates when compared to developed countries. Most emerging economies have implemented an inflation-targeting monetary policy framework in which inflation serves as the intermediate target. 
This paper outlines a novel machine learning framework, namely FEWNet, that combines the MODWT process with the ARNNx model for long-term CPI inflation forecasting for the BRIC countries under macroeconomic and geopolitical uncertainties. The proposed method demonstrates consistent performances for most of the BRIC economies for both the semi-long-run and long-run forecast horizons. Since the CPI inflation data of China presents a stationary behavior, our proposal failed to outperform all the benchmark forecasters that work exceptionally well on stationary time series. Moreover, it produces the most accurate performance for Brazil, Russia, and India, outperforming state-of-the-art machine learning or deep learning algorithms. We also present a framework for time series feature engineering using different economic filters (HP and CF band-pass filters) that have been leveraged in the proposed architecture, which supports the inclusion of exogenous regressors within its flexible design. A combination of the multiresolution analysis and autoregressive neural networks are used to predict non-linearity, non-stationarity, and non-gaussian time series of inflation. MODWT decomposition produces a trend time series and multiple detail series that contain system and slow dynamics, which easily separate signal from noise. Each time series is further modeled with ARNNx. Both theoretically and empirically, we show that under certain conditions, the proposed FEWNet reduces the empirical risk, resulting in improved forecast accuracy. 

While FEWNet is designed here to predict CPI inflation in BRIC countries, its flexible framework allows for seamless adaptation to other developed countries, including the OECD countries, and the inclusion of various additional uncertainty variables like Twitter (X)-based uncertainty measures, financial stress indicators, and broader indices like the World Uncertainty Index. This can be considered future research for this study. Moreover, the FEWNet framework is not limited to inflation forecasting alone. It holds significant potential for forecasting a wide array of macroeconomic variables like Gross Domestic Product (GDP), gold prices, unemployment rates, and more. Its strength lies in effectively handling non-stationary and non-linear time series, which are common in macroeconomic contexts. Given its capabilities, FEWNet presents itself as a robust alternative for central banks and policymakers, which often require accurate forecasts in complex economic environments. In light of our results, further analysis of the FEWNet's performance in predicting other macroeconomic time series as well as inflation in a multivariate forecasting setting seems to be a promising avenue for future research.
% Bibliography.

\section*{Acknowledgement}
The authors would like to acknowledge the Associate Editor and two experienced reviewers for insightful suggestions that improved the paper. The authors want to thank Madhurima Panja of IIIT Bangalore, India, for helping with the graphics presented in this manuscript. 

\section{Appendix}

\subsection{Discrete Wavelet Transformation:}\label{App_DWT}

The FEWNet framework leverages a form of discrete wavelet transformation (DWT) to first denoise the CPI inflation series for the BRIC countries, followed by modeling of the resulting components with the auxiliary information (trend and cycle generated through HP and CF filters) using ARNNx \cite{faraway1998time} in an ensemble setup. In particular, we focus on the `maximal overlapping' version of the DWT, popularly known as MODWT. In the literature, the DWT framework has been widely used in various applications for compressing digital images, smoothing signals \cite{percival2000wavelet, walden2001wavelet}, material science \cite{li2020comparative}, atmosphere \cite{percival1997analysis}, energy \cite{yang2021forecasting}, economics, and geophysics \cite{zhu2014modwt} among many others. We begin with a brief description of the DWT approach that forms the mathematical basis of the MODWT to be used in our proposed FEWNet framework.

The DWT approach defines a class of orthogonal wavelet transformation \cite{percival1994long}. Let $\{h_l; l = 0, 1, \ldots, L-1\}$ denote a finite length discrete high-pass (wavelet) filter such that it satisfies the unit energy assumptions and it sums up to zero, i.e., 
\begin{equation} \label{Eq_wavelet_energy}
    \sum_{l = 0}^{L-1} h_l^2 = 1 \; \text{and} \; \sum_{l = 0}^{L-1} h_l = 0. 
\end{equation}
Furthermore, the wavelet filters are also restricted to be orthogonal to even shifts. Thus mathematically we can denote this as \begin{equation}\label{Eq_even_shift}
    \sum_{l = 0}^{L-1} h_l h_{l+2\tilde{n}} = 0 \; \; \text{for any non zero integer } \tilde{n}.
\end{equation}
Alongside the high-pass filter, having a low-pass (scaling) filter $\{g_l; l = 0, 1, \ldots, L-1\}$ is also crucial for decomposing an observed time series into high-frequency oscillations and low-frequency details. These scaling filters also satisfy the unit energy assumptions and are orthogonal to even shifts (as in Eqs. (\ref{Eq_wavelet_energy}), (\ref{Eq_even_shift})). Furthermore, for all the wavelet filters, the low-pass filter coefficients are determined by the `quadrature mirror' relationship, i.e., 
$g_l = (-1)^{l+1} h_{L-1-l} \; \; \text{or} \;\; h_l = (-1)^{l} g_{L-1-l}; \; l = 0, 1, \ldots, L-1.$

\noindent For the construction of DWT coefficients from a given series, the use of the ``pyramid algorithm" is prevalent \cite{percival1994long, percival2000wavelet}. Suppose we denote the economic time series to be transformed as $Y = \{Y_t, t = 1, 2, \ldots \mathcal{N}\}$ of length $\left(\mathcal{N}=2^{\mathcal{K}}\right)$. In the pyramid algorithm, given the series $Y$, the wavelet filters $h_l$, and the scaling filters $g_l$, the first iteration performs a down-sampling operation and convolutes the data vector with each of the filters to obtain the resulting wavelet and scaling coefficient. In the subsequent step, similar down-sampling and convolution take place on the scaling coefficients of the previous step. Following this procedure iteratively the $\tilde{k}^{th}$ stage output of the pyramid algorithm can be expressed as 
\begin{equation} \label{pyramid_alg}
       \varpi_{\tilde{k},t} = \sum_{l = 0}^{L_{\tilde{k}}-1} h_{\tilde{k},l} Y_{2^{\tilde{k}}(t+1)-l \text{ mod } \mathcal{N}} \;\; \text{and} \; \; 
       v_{\tilde{k},t} = \sum_{l = 0}^{L_{\tilde{k}}-1} g_{\tilde{k},l} Y_{2^{\tilde{k}}(t+1)-l \text{ mod } \mathcal{N}}, 
\end{equation}
where $\varpi_{\tilde{k},t}$ and $v_{\tilde{k},t}$ are the $\tilde{k}^{th} \; (\tilde{k} = 1, 2, \ldots, \mathcal{K})$ level wavelet and scaling coefficients respectively. This procedure can be repeated up to $\operatorname{log}_2(\mathcal{N})$ times. Thus, using the DWT on the given time series, we can represent the discrete wavelet coefficients in matrix notation as 
\begin{equation} \label{DWT_Eqn}
    \varpi = \mathcal{W}Y,
\end{equation}
where $\mathcal{W}$, an orthogonal matrix of order $\mathcal{N} \times \mathcal{N}$, comprises of wavelet and scaling filters arranged in a row-by-row basis. The wavelet coefficients $\varpi$ can be arranged into $\mathcal{K}+1$ vectors as $\varpi = [\varpi_1, \varpi_2, \ldots, \varpi_{\mathcal{K}}, v_{\mathcal{K}}]^T$ with each of $\varpi_{\tilde{k}}, \;({\tilde{k}} = 1, 2, \ldots, \mathcal{K})$, a vector of length $\mathcal{N}/ 2^{\tilde{k}}$, consists of the wavelet coefficients associated with changes on a scale of length $2^{\tilde{k}-1}$ and the vector $v_{\mathcal{K}}$ of length $\mathcal{N}/2^{\mathcal{K}}$ is associated with averages on a scale of length $2^{\mathcal{K}-1}$. 

Although the DWT is a useful mathematical tool for decomposing discrete time series into simpler parts using high-pass and low-pass filters, it suffers from certain serious limitations. To begin with, the DWT approach down-samples the signal at each iteration, and it can be repeated for a maximum of $\operatorname{log}_2\left(\mathcal{N}\right)$, thus restricting the size of the signal to be an exact power of 2. Moreover, the wavelet and scaling coefficients generated from a DWT convoluted signal do not scale and are not shift invariant. This restricts the universal application of the DWT algorithm for arbitrary time series. To overcome these deficiencies of the DWT framework, a modified maximal overlapping version of DWT was proposed \cite{percival1994long, percival2000wavelet, walden2001wavelet}.

\subsection{MODWT approach}

The maximal overlapping discrete wavelet transform (MODWT) is an improved, non-subsampled version of the DWT framework. The MODWT overcomes the limitations of the DWT approach and is invariant to circular shifts. This non-decimated wavelet transform is suitable for handling the non-stationary behavior of a discrete-time process and is highly redundant. Moreover, the variance estimator pertaining to the MODWT process is asymptotically more efficient than that of the DWT \cite{percival1994long}. Thus, in our study, we utilize the transition and time-invariant \cite{pesquet1996time} MODWT transformation for denoising the CPI inflation series. The mathematical formulation of the MODWT algorithm can be extended from the DWT approach and is described below. 

For the MODWT algorithm, we define the rescaled version of filters as $\tilde{h}_l  = h_l/2^l$ and $\tilde{g}_l = g_l/2^l$ and employ the pyramid algorithm with the time series $Y$, wavelet filter $\tilde{h}_l$, and scaling filter $\tilde{g}_l$. Similar to the DWT approach in the non-decimated MODWT framework, the pyramid algorithm convolutes the data with the re-normalized filters and generates the resulting coefficients of length $\mathcal{N}$ as 
\begin{equation} 
       \tilde{\varpi}_{\tilde{k},t} = \sum_{l = 0}^{L_{\tilde{k}}-1} \tilde{h}_{\tilde{k},l} Y_{(t+1)-l \text{ mod } \mathcal{N}} \;\; \text{and} \; \; 
       \tilde{v}_{\tilde{k},t} = \sum_{l = 0}^{L_{\tilde{k}}-1} \tilde{g}_{\tilde{k},l} Y_{(t+1)-l \text{ mod } \mathcal{N}}, 
\end{equation}
where $L_{\tilde{k}} = (2^{\tilde{k}} - 1)(L-1)+1$. The wavelet ($\tilde{\varpi}_{\tilde{k},t}$) and scaling ($\tilde{v}_{\tilde{k},t}$) coefficients obtained by decomposing the data vector $Y$ using the MODWT approach into $\mathcal{K}$ levels can be formulated as 
\begin{equation} \label{MODWT_PA}
    \tilde{\varpi} = \tilde{\mathcal{W}} Y
\end{equation}
where the orthonormal matrix $\tilde{\mathcal{W}} = [\tilde{\mathcal{W}}_1, \tilde{\mathcal{W}}_2, \ldots, \tilde{\mathcal{W}}_{\mathcal{K}}, \tilde{V}_{\mathcal{K}}]^T$ is made up of $\mathcal{K}+1$ submatrices of size $\mathcal{N} \times \mathcal{N}$ with $\tilde{\mathcal{W}}_{\tilde{k}}, \; (\tilde{k} = 1, 2, \ldots, \mathcal{K})$ and $\tilde{V}_{\mathcal{K}}$ denoting circularly shifted wavelet and scaling filters respectively. Furthermore, the wavelet and scaling coefficients can be arranged in $\mathcal{K}+1$ vectors as $\tilde{\varpi} = [\tilde{\varpi}_1, \tilde{\varpi}_2, \ldots, \tilde{\varpi}_{\mathcal{K}}, \tilde{v}_{\mathcal{K}}]^T$ in a similar manner as in Eq. (\ref{DWT_Eqn}) of DWT approach.

Following the MODWT decomposition, an additive multi-resolution analysis (MRA) of the observed time series $Y$ can be formulated as 
\begin{equation}
    Y_t = \sum_{\tilde{k} = 1}^{\mathcal{K}} d_{\tilde{k}, t},
\end{equation}
where $d_{\tilde{k}, t}, \; t = 1, 2, \ldots, \mathcal{N}$ is the $t^{th}$ element of $d_{\tilde{k}} = \tilde{\mathcal{W}}_{\tilde{k}} \tilde{\varpi}_{\tilde{k}}, \; ({\tilde{k}} = 1, 2, \ldots, \mathcal{K})$. Thus, the time series $Y$ can be expressed as a linear combination of the high-frequency and low-frequency coefficients as 
\begin{equation}
    Y_t = S_{\mathcal{K},t} + \sum_{\tilde{k} = 1}^{\mathcal{K}} \delta_{\tilde{k},t}, 
\end{equation}
where $S_{\mathcal{K},t} = \sum_{j = \tilde{k}+1}^{\mathcal{K}+1} d_{j, t}$ is the $\mathcal{K}^{th}$ level smooth coefficient and $\delta_{\tilde{k},t} = \sum_{j = 1}^{\tilde{k}} d_{j, t}$ is the $\tilde{k}^{th}$ level details associated with the MODWT-based MRA decomposition. A salient feature of this scale-based additive transformation is that the wavelet and scaling coefficients of the series are associated with the zero-phase filters. Hence, they are of the same size as the original series and are perfectly aligned with it. 
A more detailed structural overview of the MODWT process can be found in \cite{percival1997analysis}. 

\subsection{Economic filters - Hodrick-Prescott and Christiano-Fitzgerald filters:}\label{App_Exo_fil}
The Hodrick-Prescott filter is a popular econometric method that has found profound applications in extracting the business cycle components and as a detrending method for financial time series \cite{hodrick1997postwar}. The method is named after its proponents: - Robert J. Hodrick and Edward C. Prescott. This nonparametric method is widely adopted by economists in central banks, international economic agencies, industry, and in government sector \cite{phillips2019boosting}. There is a wide array of time series filters commonly used in macroeconomic and financial research and analysis to decompose the overall time series into trend, cycle, and irregular components \cite{genccay2001introduction}. Among them, the HP filter has been in vogue among researchers for extracting trend and cycle components from the time series. One key aspect of the HP filter is that it is quite effective and can be applied to non-stationary time series. Therefore, it makes an excellent choice for decomposing CPI inflation, EPU, and GPRC series in this exercise.

For any time series data (\(Y_{t}:t = 1,\ldots.,\ n\)), the HP filter decomposes the series into a trend (\(\tau_{t}\)), and a cyclical component (\(C_{t}\)), by solving the following minimization problem with respect to \(\tau_{t}\):
\begin{equation}
    {\widehat{\tau}}_{t}^{HP} = \underset{\tau_{t}}{\operatorname{argmin}}
\left\{ \sum_{t = 1}^{n}{({Y_{t} - \tau_{t})}^{2}} + \ \theta\sum_{t = 2}^{n}{({\mathrm{\Delta}^{2}\tau_{t})}^{2}} \right\} \; \text{and} \;
{\widehat{C}}_{t}^{HP} =
Y_{t} - {\widehat{\tau}}_{t}^{HP} 
\end{equation}
where \(\mathrm{\Delta}\tau_{t}\) = \(\tau_{t}\) - \(\tau_{t - 1}\),
\(\mathrm{\Delta}^{2}\tau_{t}\) =
\(\mathrm{\Delta}\tau_{t} - \ \mathrm{\Delta}\tau_{t - 1}\) =
\(\tau_{t} - 2\tau_{t - 1}\) + \(\tau_{t - 2}\), and  \(\theta \geq 0\) is a tunable parameter that controls the extent of penalty for fluctuations in the second differences of \(Y_{t} \)or in other words, it controls for the volatility in the trend component. The greater the value of \(\theta,\), the more stable the trend component is. The choice of \(\theta\) is determined by the user. Quite interestingly, the residual term or deviation from the trend term: (\(Y_{t} - \ {\widehat{\tau}}_{t}^{HP}\)) is defined as the business cycle components. Thus, the HP filter can be considered as a \say{highpass filter}, removing the trend components and returning high-frequency components in \({\widehat{C}}_{t}^{HP}\). Let's assume that a time series be expressed as: \(Y_{t} = \ \tau_{t}\)+\(C_{t}\) (where \(C_{t}\) incorporates the effect of any irregular term). If \(C_{t}\) and the 2\textsuperscript{nd} difference of \(\tau_{t}\) are normally and independently distributed, the HP filter would be an \say{optimal filter} and \(\theta\) is given as the ratio of the two variances:
\(\frac{\sigma_{C}^{2}}{\sigma_{\mathrm{\Delta}_{\tau}^{2}}^{2}}\),
where \(\mathrm{\Delta}^{2}\) is the 2\textsuperscript{nd} difference operator as defined earlier. One very interesting property of \(\theta\) is that, in the limiting case, when \(\theta \rightarrow \infty\), the trend component becomes a linear time trend. In common practice, \(\theta\) =1600 is used, as prescribed by \cite{hodrick1997postwar}, when the HP filter is applied to quarterly US economic data. For other frequencies like annual or monthly data, \(\theta\) =6.25 and \(\theta\) =129,600 are used as default values \cite{ravn2002adjusting}. The cyclical or residual component derived from the HP filters have the following frequency response function \cite{genccay2001introduction}:
\begin{equation}
\mathcal{H}(f,\theta) =  \frac{4\theta\lbrack 1 - cos(2\pi f{)\rbrack}^{2}}{1 + 4\theta\lbrack 1 - \cos(2\pi f)\rbrack^{2}} 
\end{equation}
and the trend component of the HP filter is given by:
\begin{equation}
\tau_{t}= \frac{\phi_{1}\phi_{2}}{\theta} \left[\sum_{j = 0}^{\infty}{(A_{1}\phi_{1}^{j} + A_{2}\phi_{2}^{j})Y_{t - j}} + \sum_{j = 0}^{\infty}{(A_{1}\phi_{1}^{j} + A_{2}\phi_{2}^{j})Y_{t + j}}\right],
\end{equation}
where \(\phi_{1}\ {and\ \phi}_{2}\) are the complex conjugates and depends on the value of \(\theta\ \)and \(A_{1}\) and \(A_{2}\) are the functions of \(\phi_{1}\ and\ \phi_{2}\). By this analysis, it can be argued that the trend component, extracted using the HP filter, is a centered moving average term.

On the other hand, the Christiano-Fitzgerald (CF) (often regarded as a random walk filter) filter is a band pass filter that was primarily constructed using very similar principles as Baxter and King (BK) filter \cite{christiano2003band}. CF filter has demonstrated better performance compared to the BK filter in the case of real-time applications. CF filter can be calculated as:
\begin{equation}
C_{t}=\varphi_{0} Y_{t}+\varphi_{1} Y_{t+1}+\cdots+\varphi_{T-1-t} Y_{T-1}+\tilde{\varphi}_{T-t} Y_{T}+\varphi_{1} Y_{t-1}+\cdots+\varphi_{t-2} Y_{2}+\tilde{\varphi}_{t-1} Y_{1}
\end{equation}
where $\varphi_{j}=\frac{\sin (j b)-\sin (j a)}{\pi j}, j \geq 1$, and $\varphi_{0}=\frac{b-a}{\pi}, a=\frac{2 \pi}{p_{u}}, b=\frac{2 \pi}{p_{l}}$ and $\tilde{\varphi}_{k}=-\frac{1}{2} \varphi_{0}-\sum_{j=1}^{k-1} \varphi_{j}$. The parameters, $p_{u}$ and $p_{l}$ represent the break-off point of cycle length in the month. Any cycle value between $p_{u}$ and $p_{l}$ is preserved in the cyclical term $C_{t}$. CF filter aims to formulate the de-trending and smoothing problem in the frequency domain. The CF filter approximates the ideal infinite bandpass filter, which is like the BK filter. One key advantage of CF filters is that the random walk filter can exploit the complete series for the calculation of each filtered point. CF filters can converge well with optimal filters in the long run. This property speaks for the effectiveness of the CF filter as compared to the BK filter in the context of extracting meaningful features (cyclical components) from a specific time series.

In this analysis, the HP filter has been explored to extract the trend component for CPI inflation, EPU, and GPRC series Whereas, the CF filter is used to extract the cyclical terms from all the 3-time series (CPI inflation, EPU, and GPRC) and post that these derived features are used in the downstream FEWNet algorithm to generate the long-term forecasts for the CPI inflation series.

\subsection{Proof for Proposition \ref{prop_1}}\label{Prop_proof}
%\begin{proof}
\noindent\textbf{Proof.}
Using Eq. (\ref{Eq_new_star}), we write the ARNNx estimator as follows
\begin{equation}    
    \hat{f} = \alpha + \sum_{j^{'} = 1}^{q^{*}} \beta_{j^{'}} \sigma\left( z_{j^{'}}\right),
\end{equation}
where $z_{j^{'}} = \alpha^{*} + \sum_{i = 1}^r \beta_{i, j^{'}}^{*} y_i + \gamma^{*}\underbar{X}$ where the parameters are estimated by minimizing Eq. (\ref{New_eq_111}). For FEWNet, we have
\begin{equation}
\begin{gathered}
    \hat{f}_0 = \alpha_{0}^{S} + \sum_{i = 1}^q \beta_{i}^{S} \sigma(\alpha_{i}^S + \sum_{j_0 = 1}^{r}{\beta^{S}_{i, j_0}} S_{\mathcal{K},j_0} + {\gamma^{S}_{i}}^{'} \underbar{X})\\
    \hat{f}_{\tilde{k}} = \alpha_{0, \tilde{k}} + \sum_{i = 1}^q \beta_{i, \tilde{k}} \sigma(\alpha_{i, \tilde{k}} + \sum_{j_{\tilde{k}} = 1} ^ r \beta_{i, j_{\tilde{k}}} {\delta}_{\tilde{k}, j_{\tilde{k}}} + \gamma^{'}_{i, \tilde{k}} \underbar{X}); \; \tilde{k} = 1, 2, \ldots, \mathcal{K}.
\end{gathered} 
\end{equation}
Using Jensen's inequality and Eq. (\ref{Eq_ERM_3}), we get the $\mathcal{R}_{\text{Emp}}$ definition domain is a subspace of $\mathcal{R}_{\text{Emp}}^{W}$
\begin{equation}
    \operatorname{min} \mathcal{R}_{\text{Emp}}^W \leq \operatorname{min} \mathcal{R}_{\text{Emp}}
\end{equation}
Since $\hat{f}$ is a linear combination of $\hat{f}_0$ and $\hat{f}_{\tilde{k}}, \; (\tilde{k} = 1, 2, \ldots, \mathcal{K})$ with uniform weights and scaling ($\tilde{g}_l$) and wavelet ($\tilde{h}_l$) filter satisfies unit energy assumption and even length scaling assumption in MODWT i.e., 
\begin{equation} \label{Assumptions}
    \tilde{h}_l = - \tilde{g}_l \; \forall l \geq 1, \;\; \tilde{h}_0 + \tilde{g}_0 = 1, \; \; \tilde{h}_l = \tilde{g}_l = 0 \; \forall l \leq -1
\end{equation}
Similar results may hold for any other decomposition method satisfying the Eq. (\ref{Assumptions}) in fact. 
%\end{proof}
\subsection{Conformal Prediction for semi-long-term}\label{App_CP_SL}

\begin{figure}[!ht]
 \centering
  \includegraphics[width=0.9\textwidth]{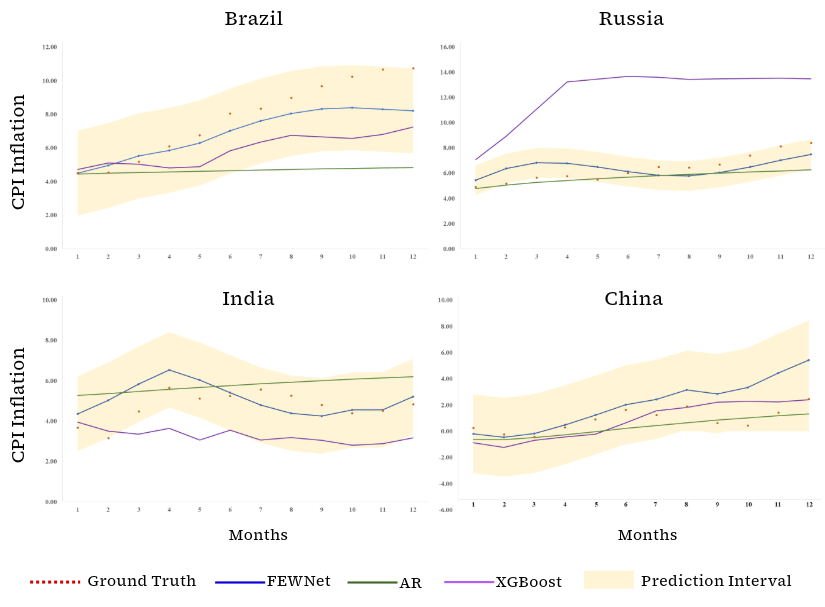}
   \caption{The plot shows the ground truth CPI inflation data (red dots), 12 months point forecasts generated by FEWNet (blue line), AR (green line), XGBoost (purple line), and the conformal prediction interval of FEWNet (yellow shaded) for BRIC Countries}
   \label{fig:CP_Semi_long}
\end{figure}

Figure \ref{fig:CP_Semi_long} depicts the uncertainty quantification of the FEWNet framework for the 12 months ahead forecast horizon. The conformal prediction intervals demonstrated in the figure are of varied widths across countries. In the case of 12 months forecasts for Brazil, the width of the interval averages around 5.06, while for Russia, it hovers around 2.3, for India, it averages 3.7, and the width value for China is around 6.03, the maximum among all the countries. This indicates a higher degree of prediction uncertainty for China than for other countries. 

\bibliography{refs}

\end{document}